\documentclass[showpacs,aps,twocolumn]{revtex4-1}

\usepackage{bm}
\usepackage{amsmath}
\usepackage{graphicx}
\usepackage{subfigure}
\usepackage[usenames,dvipsnames]{color}
\definecolor{darkblue}{RGB}{0,0,196}
\usepackage[colorlinks=true,linkcolor=darkblue,citecolor=darkblue,urlcolor=darkblue]{hyperref}
\usepackage{setspace}
\usepackage{footmisc}
\usepackage[makeroom]{cancel}
\usepackage{multirow}

\usepackage{color,soul}

\def\be{\begin{equation}}
\def\ee{\end{equation}}
\def\ba{\begin{eqnarray}}
\def\ea{\end{eqnarray}}

\usepackage{graphicx}
\usepackage{amsmath,bbm}
\usepackage{amssymb,bm}

\begin{document}

\title{Exploring Anisotropic flow via the Boltzmann Transport Equation Employing the Tsallis Blast Wave Description at LHC energies}
\author{Aviral Akhil}
\author{Swatantra~Kumar~Tiwari}
\email{sktiwari4bhu@gmail.com (Corresponding Author)}
\affiliation{Department of Physics, University of Allahabad, Prayagraj- 211002, U.P.}

\begin{abstract}
\noindent
Anisotropic flows $i.e.$ azimuthal anisotropies in the particle production are one of the important probes in characterizing the properties of the strongly interacting matter created in the relativistic heavy-ion collisions. These observables are sensitive to both the transport properties as well as the equation of state (EOS) of the Quantum Chromodynamics (QCD) matter. We have adopted the Boltzmann transport equation (BTE) in the relaxation time approximation (RTA) to describe the  experimental data for harmonic flows such as elliptic flow ($v_2$), triangular flow ($v_3$), quadrangular flow ($v_4$) obtained in heavy- ion collisions at Large Hadron Collider (LHC) energies. In this analysis, we have used Tsallis statistics as an initial distribution and the Tsallis Blast wave (TBW) description is used as the equilibrium distribution function while describing the evolution of the particle production in BTE. We have fitted the transverse momentum spectra, $v_2$, $v_3$, and $v_4$ of identified hadrons such as pion, kaon, and proton for Pb-Pb and Xe-Xe collisions at the LHC energies of $\sqrt{s_{NN}}$ = 5.02 TeV and $\sqrt{s_{NN}}$ = 5.44 TeV, respectively for various centralities. Our study offers the comparative analysis between the two distinct collision systems operating at comparable collision energies. The present formulation successfully fits the experimental data for $p_T$- spectra upto $p_T$ = 8 GeV and effectively explains the anisotropic flows data upto $p_T$ = 10 GeV with a very favourable $\chi^2/ndf$. We observe that the average transverse flow velocity ($<\beta>$) and the kinetic freeze-out temperature ($T$) extracted in our analysis decrease as we go towards the peripheral collisions. Non-extensive parameters ($q_{AA}$ and $q_{pp}$) exhibit an ascending trend from central to peripheral collisions, signifying an almost thermalized system in the most central collisions and a non-equilibrium state in peripheral ones. The azimuthal modulation amplitudes ($\rho_a$) for $v_2$, $v_3$, and $v_4$ exhibit an increasing pattern as one moves from the most central to peripheral collisions in both the Pb-Pb and Xe-Xe nuclei interactions.

\end{abstract}


\date{\today}

\maketitle 
\section{Introduction}
\label{intro}

The investigation of high-energy heavy ion collisions has emerged as a cornerstone of modern nuclear and particle physics, offering a unique window into the fundamental properties of matter under extreme conditions. A key aspect of these collisions is the intricate interplay between the participating particles, leading to complex momentum transfer phenomena that shape the evolution of the collision dynamics. The quantitative understanding of these momentum transfer is pivotal not only for unraveling the underlying physics but also for informing the development of advanced theoretical models and experimental strategies. One of the prime goals of relativistic heavy-ion collision programs is to characterize the properties of the hot and dense medium known as Quark Gluon Plasma (QGP) created in these collisions. Earlier investigations conducted at the Super Proton Synchrotron (SPS), CERN~\cite{Heinz:2000bk}, at the Relativistic Heavy Ion Collider (RHIC)~\cite{BRAHMS:2004adc,PHOBOS:2004zne,PHENIX:2004vcz,STAR:2005gfr, Shuryak:2004cy} and at the Large Hadron Collider (LHC)~\cite{ALICE:2022wpn, Schukraft:2011np,Steinberg:2011dj,Wyslouch:2011zz} have yielded compelling evidence suggesting the presence of the QGP, setting the stage for further in-depth exploration in heavy ion collision experiments. The properties of this medium can be studied via azimuthal anisotropies of the produced particles in the momentum space. These momentum anisotropies are arise due to the initial state geometry asymmetries. These asymmetries are characterized by the Fourier expansion coefficients $v_2$, $v_3$, $v_4$, etc., of the azimuthal distribution of the particles. Experimentally, the azimuthal anisotropies for the hadrons created in the heavy- ion collisions have been studied at the RHIC \cite{STAR:2002hbo,STAR:2004jwm, PHENIX:2018hho, PHOBOS:2004vcu} and at the LHC~\cite{ALICE:2010suc,ALICE:2011ab} energies. Theoretically, relativistic hydrodynamics are extensively used to study the anisotropic flows measured in the heavy-ion collisions for the recent review on hydrodynamics see ref.~\cite{Jeon:2015dfa}).

The Boltzmann Transport Equation (BTE), a fundamental concept in the statistical mechanics, finds a profound application in the study of the medium created in the heavy ion collisions. Emerging from the kinetic theory of gases, it provides a mathematical framework for understanding the intricate dynamics of particles within the extreme conditions generated during high-energy heavy- ion collisions. In the context of the heavy ion collisions, this equation serves several critical purposes. The equation can be adapted to study collective phenomena, such as the development of flow patterns within the medium. This helps us in understanding how the initial state of the colliding nuclei evolves into a complex, collective behaviour, shedding light on the properties of the created medium. In this investigation, we have employed Tsallis statistics as the initial distribution, and for elucidating the particle production evolution within the Boltzmann Transport Equation (BTE), we have adopted the Tsallis Blast Wave (TBW) description as the equilibrium distribution function.

Boltzmann-Gibbs statistics, which underlie classical thermodynamics, assumes that systems in equilibrium are described by the exponential distribution and relies heavily on the concept of entropy. However, in some complex systems, such as those with long-range interactions, fractal structures, or in non-extensive thermodynamics, Boltzmann-Gibbs statistics may not be adequate. Tsallis statistics~\cite{Tsallis:1987eu} introduces a modified form of entropy, now called the Tsallis entropy ($S_q$), which is parametrized by a non-extensive parameter,$"q"$. The Tsallis entropy leads to a generalized probability distribution function, known as the Tsallis distribution (or Tsallis q-distribution). This distribution plays a crucial role in the study of complex systems and has found applications in the various fields of physics, including the exploration of the properties of QGP~\cite{De:2007zza,Wilk:2008ue,Alberico:1999nh,Biro:2003vz,Osada:2008sw} and the improvement of the Boltzmann transport equation.

Boltzmann-Gibbs blast- wave (BGBW) model~\cite{Chen:2020zuw, Song:2019sez} has long served as a fundamental pillar in this field. The BGBW model, a stalwart in this domain, postulates a critical assumption: the system reaches a local thermal equilibrium at a specific moment in time before embarking on a hydrodynamic evolution~\cite{Che:2020fbz}. It has successfully described observables such as transverse momentum distributions of identified particles, offering valuable insights into the transverse expansion and the temperature at the moment when hadrons decouple from the system. In references~\cite{Jaiswal:2015saa,Yang:2016rnw}, the viscous blast-wave model is used to study the transverse momentum spectra and azimuthal anisotropies of various particles measured in heavy-ion collision experiments. In a recent work~\cite{Moriggi:2023ahi}, the origin of azimuthal asymmetry in nuclear collisions is explored by employing a model that considers both the particles generated in the initial high-energy collisions and the collective effects characterized by an expansion similar to a Blast-Wave. However, the BGBW model faces a significant challenge, the inherent fluctuations in initial conditions~\cite{Socolowski:2004hw}, which fluctuate unpredictably from one collision event to another. These fluctuations can profoundly influence particle spectra, particularly in the low and intermediate transverse momentum ($p_T$) range. To account for the influence of fluctuations, the authors~\cite{Tang:2008ud} opted to replace the Boltzmann distribution with the Tsallis distribution~\cite{Tsallis:1987eu} for modelling the particle emission source, thereby adapting the statistical framework to better accommodate fluctuation-related phenomena. This fundamental shift has given rise to the Tsallis blast-wave (TBW) model, uniquely equipped to explore particle spectra in details. TBW model has been employed to scrutinize the spectra of a variety of particles, encompassing $\pi^{\pm}$, $k^{\pm}$, $p(\bar{p})$, $\phi$, $\Lambda(\bar\Lambda)$, and $\Xi^-(\Xi^+)$ for Au-Au collisions at $\sqrt{s_{NN}}$ = 200 GeV~\cite{Tang:2008ud}. In ref.~\cite{Shao:2009mu}, TBW was undertaken to encompass the spectra of both strange and non-strange hadrons. The results revealed a notable distinction in the central collisions, where strange hadrons exhibited smaller non-extensive parameter and average transverse flow velocity values alongside higher temperatures compared to the non-strange hadrons. This observation implies a potential earlier decoupling of strange hadrons relative to non-strange ones. In ref.~\cite{Che:2020fbz}, the authors delved into the particle spectra of Pb-Pb, Xe-Xe, and p-Pb collisions at energies of 2.76 TeV (for Pb-Pb), 5.02 TeV (for Pb-Pb and p-Pb) and 5.44 TeV (for Xe-Xe) using the TBW model, incorporating both linear and constant velocity profiles. Generally, the model successfully captures the spectra upto $p_T$ = 3 GeV. Notably, they observed that as collisions transition from central to peripheral, average transverse flow velocity decreases, whereas temperature and non-extensive parameter exhibit the opposite trend. This suggests that in more central collisions, the system experiences a more rapid expansion and maintains a lower degree of off-equilibrium behaviour. In ref.~\cite{Che:2020fbz}, TBW is used to analyse the $p_T$- spectra of identified particles at $\sqrt{s_{NN}}$ = 2.76 TeV and 5.02 TeV for Pb- Pb collisions, $\sqrt{s_{NN}}$ = 5.02 TeV for p-Pb collisions as well as for Xe- Xe collisions at $\sqrt{s_{NN}}$ = 5.44 TeV. They successfully describe the experimental data upto $p_T$ = 3 GeV with a very good $\chi^2/ndf$.  

The manuscript is structured as follows: Section~\ref{formulation} presents the derivation of transverse momentum spectra and azimuthal anisotropies, utilizing the Boltzmann Transport Equation in the relaxation time approximation. In Section~\ref{RD}, we delve into a comprehensive discussion of the results obtained. Lastly, in Section~\ref{summary}, we provide a succinct summary of the study along with potential future directions.

\begin{figure*}[htb]
\subfigure[\ Transverse momentum spectra of pion, kaon and proton for 0-5 $\%$ centrality in Pb-Pb collisions.]{
\begin{minipage}[b]{0.48\textwidth}
\centering \includegraphics[width=\linewidth]{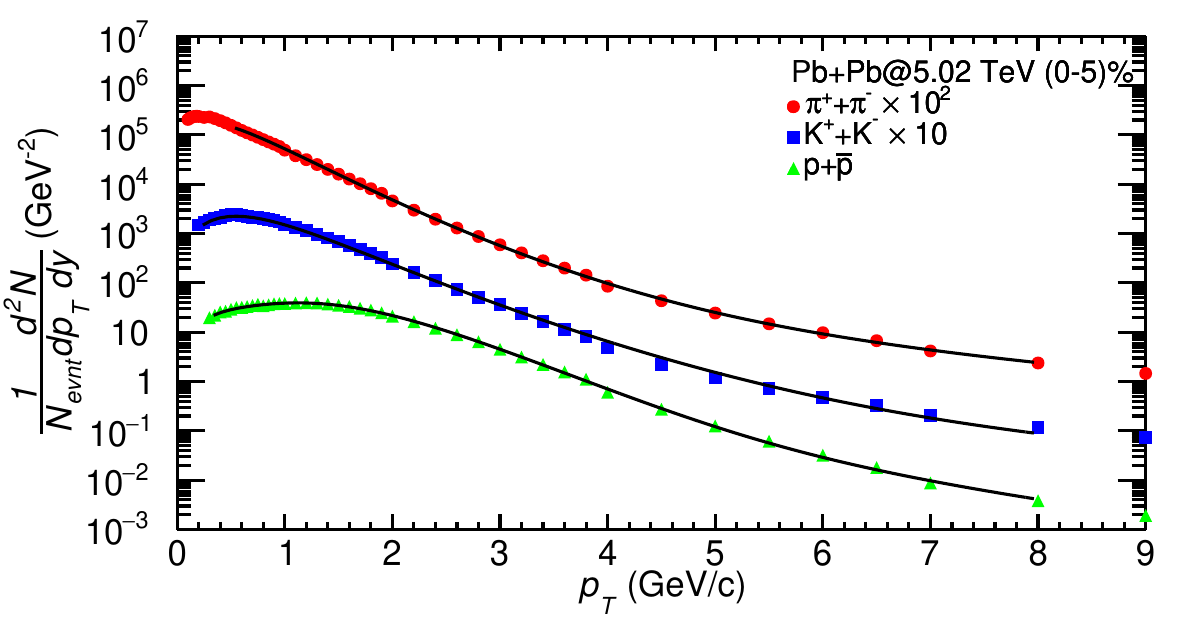}
\end{minipage}}
\hfill
\subfigure[Transverse momentum spectra of pion, kaon and proton for 70- 80 $\%$ centrality in Pb-Pb collisions.]{
\begin{minipage}[b]{0.48\textwidth}
\centering \includegraphics[width=\linewidth]{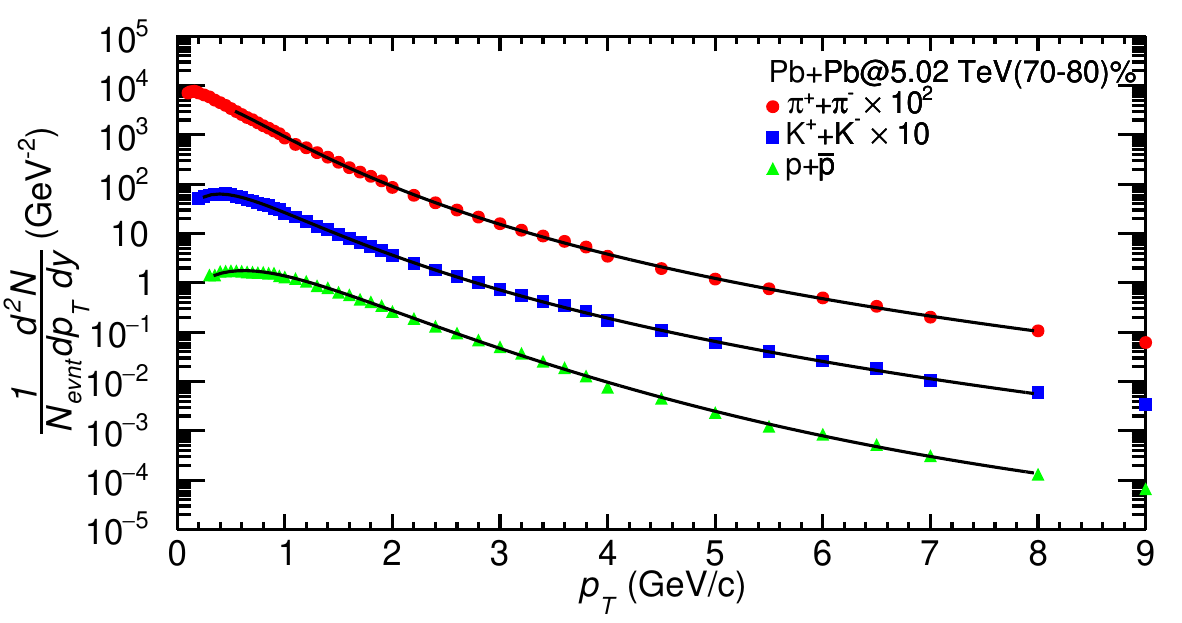}
\end{minipage}}
\hfill
\subfigure[\ Transverse momentum spectra of pion, kaon and proton for 0-5 $\%$ centrality in Xe-Xe collisions.]{
\begin{minipage}[b]{0.48\textwidth}
\centering \includegraphics[width=\linewidth]{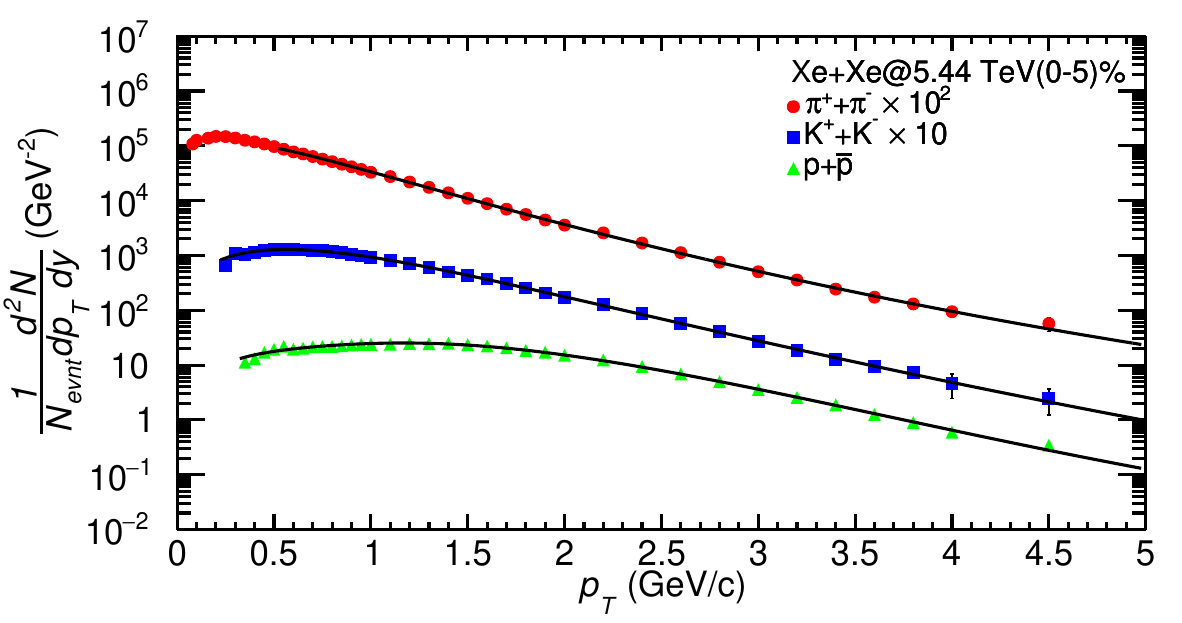}
\end{minipage}}
\hfill
\subfigure[\ Transverse momentum spectra of pion, kaon and proton for 60-70 $\%$ centrality in Xe-Xe collisions.]{
\begin{minipage}[b]{0.48\textwidth}
\centering \includegraphics[width=\linewidth]{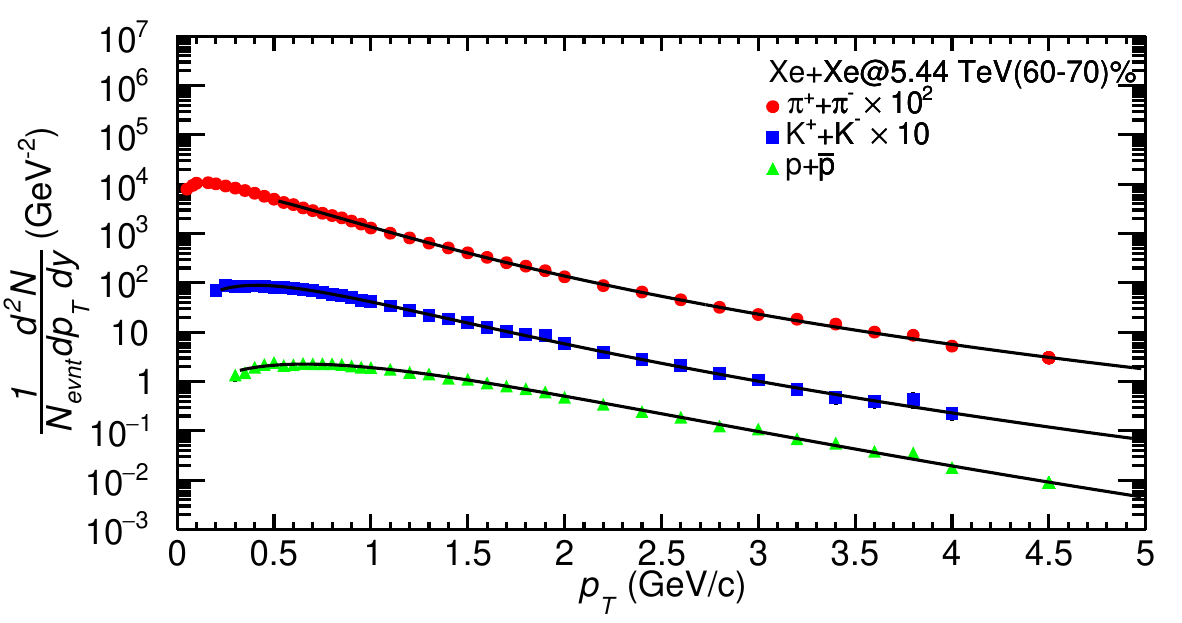}
\end{minipage}}
\caption{The transverse momentum spectra of identified hadrons in Pb-Pb and Xe-Xe collisions for the most central and peripheral collisions.}
\label{fig1} 
\end{figure*}

\begin{figure}
\includegraphics[height=12em]{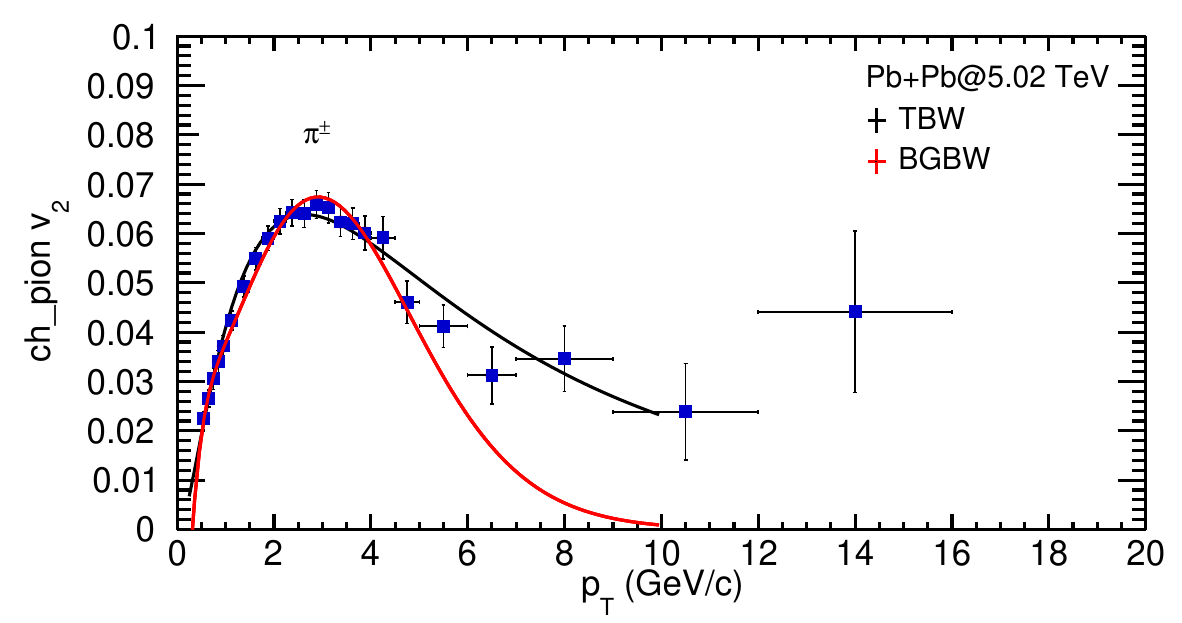}
\caption[]{(colour online) The comparison between the BGBW and TBW used as $f_{eq}$ in BTE with RTA.}
\label{comp}
\end{figure}

\section{Formulation}
\label{formulation}

The particle distribution in the four-momentum space can be written as a Fourier series~\cite{Sun:2014rda},

\begin{equation}
 \label{eq1}
  E\frac{d^3N}{dp^3}=\frac{1}{2\pi}\frac{d^{2}N}{p_{T}dp_{T}dy}\left(1+2\sum_{n^\prime=1}^{\infty} v_{n^\prime}\,\cos(n^{\prime}\phi)\right),
\end{equation}

where $E$ represents the energy of the emitted particles, which is a crucial parameter characterizing the particles$'$ properties. $y$ denotes the rapidity of these particles, a relativistic measure of their momentum along the beam direction,  $\phi$ is the azimuthal angle of a particle and $v_{n^\prime}$ is the $n^{\prime th}$ harmonic coefficients that decode it's motion's patterns as: $v_2$ unveils elliptic flow, $v_ 3$  captures triangular correlations, and $v_4$ reveals quadrangular patterns and so on. These coefficients hold clues to early pressure gradients, medium viscosity, and particle interactions, unveiling the complexities of the collision process.

Previously, various theoretical computations grounded in transport equations~\cite{Kolb:2000fha, Das:2017dsh} and phenomenological models~\cite{Younus:2012yi,Tripathy:2017nmo} had successfully addressed the explanation of $v_2$. However, in this current study, we embark on an innovative endeavour, elucidating not only $v_2$ but also $v_3$ and $v_4$ through the utilization of the Tsallis Blast wave description within the framework of the Boltzmann Transport Equation (BTE). Subsequent sections will intricately explore the application of BTE in tracking the dynamic evolution of the particle momentum distribution within a thermodynamic milieu, leading to a comprehensive exposition of the outcomes of this article.

\noindent
\subsection{Anisotropic flows in Boltzmann transport equation (BTE) using Relaxation time approximation (RTA)}
\label{v_n}

In the arena of high-energy heavy ion collisions, the study of anisotropic flow offers insights into the complex interplay of particles amidst extreme conditions as described in the introductory section. These asymmetric flow patterns, mirroring the collision's initial geometry and subsequent dynamics, hold keys to understanding the transport properties of the medium created in the heavy-ion collision. Navigating this intricate landscape, the Relaxation Time Approximation (RTA) within the venerable Boltzmann Transport Equation (BTE) emerges as a powerful tool.

The use of RTA in BTE simplifies intricate interactions through a relaxation time parameter. This partnership unveils how particles approach equilibrium after interactions. The BTE- RTA duo shines a light on the interplay of scattering, relaxation, and viscosity. This mathematical union not only aids the interpretation of experimental results but also unravels transport coefficients' significance. This endeavour delves into RTA within BTE, focusing on anisotropic flow. We decipher intricate relationships geometry, relaxation, and flow emergence through theoretical discourse and analytical tools. Our aim is to deepen the understanding of the medium formed in heavy-ion collisions, refine theory, and guide experimental inquiry.

The BTE in general can be written as:

\begin{eqnarray}
\label{eq2}
 \frac{df(x,p,t)}{dt}=\frac{\partial f}{\partial t}+\vec{v}.\nabla_x
f+\vec{F}.\nabla_p
f=C[f]
\end{eqnarray}

The distribution of particles denoted as $f(x, p, t)$ depends upon the position, momentum, and time. Here, $\vec{v}$ represents velocity, while $\vec{F}$ stands as the external force. The notations $\nabla_x$ and $\nabla_p$ denote partial derivatives concerning position and momentum, respectively. The term $C[f]$ embodies collision interactions between the probing particles and the medium. Previously, the Boltzmann Transport Equation (BTE) within the Relaxation Time Approximation (RTA) framework has been employed to investigate various phenomena. These include the temporal progression of temperature fluctuations in non-equilibrium systems~\cite{Bhattacharyya:2015nwa}, analysis of elliptic flow of identified hadrons~\cite{Tripathy:2017nmo,Younus:2018mrk} as well as the assessment of $R_{AA}$ for diverse light and heavy flavors, at energies pertinent to the Large Hadron Collider (LHC)~\cite{Tripathy:2016hlg}.

For the sake of simplification, assuming homogeneity of the system ($\nabla_x f=0$) and in the absence of external forces ($\vec{F}=$0), the second and third terms of the Eq.~\ref{eq2} become zero and it reduces to,

\begin{equation}
 \label{eq3}
  \frac{df(x,p,t)}{dt}=\frac{\partial f}{\partial t}=C[f].
\end{equation}

In RTA \cite{Florkowski:2016qig}, the collision term is expressed as:
\begin{equation}
 C[f] =-\frac{f-f_{eq}}{\tau}
 \label{CT}
\end{equation}

where $f_{eq}$ is the Boltzmann local equilibrium distribution characterized by a temperature $T$. $\tau$ is the relaxation time, the time taken by a non-equilibrium system to reach equilibrium. 


Solving eq.~\ref{eq3} considering RTA with the initial conditions {\it i.e.} at $t=0, f=f_{in}$ and at $t=t_f, f=f_{fin}$ and taking the assumption that the equilibrium distribution function and relaxation time do not depend on time we get,
\begin{equation}
 \label{eq6}
 f_{fin}=f_{eq}+(f_{in}-f_{eq})e^{-\frac{t_f}{\tau}},
\end{equation}
where $t_f$ is the freeze-out time. Using Eq.~\ref{eq6}, the expression of the anisotropic flows ($v_{n^\prime}$) can be written as,
\begin{equation}
\label{anisotropic_flow}
v_{n^\prime}(p_T)=\frac{\int{f_{fin} \times \cos(n^{\prime}\phi)\,d\phi}}{\int{f_{fin}\,d\phi}}.
\end{equation}
Eq.~\ref{anisotropic_flow} gives the $n^{\prime th}$ azimuthal anisotropies after incorporating RTA in BTE. It involves the Tsallis non-extensive distribution function as the initial distribution of particles and TBW function as the equilibrium distribution. Continuing our discourse, we delve into the comprehensive derivation of the TBW model as done in~\cite{Che:2020fbz}. 
 In the TBW model, the invariant distribution function for identified particles is
given by~\cite{Tang:2008ud}, 
\begin{equation}
\label{bgbw1}
f_{eq}(x,p)=  \frac{g}{(2\pi)^3}\Big(1\;+(q_{AA}-1)\frac{E-\mu}{T}\Big)^\frac{-1}{q-1}.
\end{equation}

Here, the temperature denoted by $T$ is a kinetic freeze-out temperature, while $g$ signifies the degeneracy factor. The energy of the emitted particles, described by $E=p^\nu u_\nu$, originates from a source in motion with the velocity $u_\nu$ and momentum $p^\nu$. The latter can be expressed as, $p^\nu = (m_T\cosh{y}, p_T\cos{\phi_p}, p_T\sin{\phi_p}, m_T\sinh{y})$. The velocity of the source is denoted by  $u^\mu = \cosh{\rho}(\cosh{y_s}, \tanh{\rho}\cos{\phi_b}, \tanh{\rho}\sin{\phi_b}, \sinh{y_s})$, where $y$ and $m_T$ symbolize the rapidity and transverse mass of the identified particles. $y_s$  represents the rapidity of the emitting source, while $\phi_p$ and $\phi_b$ are the azimuthal angles of the emitted particle velocity and the flow velocity with respect to the x-axis in the reaction plane. The azimuthal direction of the boost,  $\phi_b$, is aligned with the azimuthal angle of the emitting source in coordinate space. The parameter, $q_{AA}$ encapsulates non-extensivity, quantifying the extent of deviation from equilibrium. This departure from unity is indicative of the non-equilibrium nature of the system. The parameter $\rho$ known as the transverse expansion rapidity~\cite{Che:2020fbz} is expressed as,  $\rho=tanh^{-1}\beta+\rho_a \cos(n^{\prime}\phi)$~ \cite{Huovinen:2001cy} where $\rho_a$  stands for the azimuthal modulation amplitude in the flow, and $\beta=\displaystyle\beta_s\;\Big(\xi\Big)^n$ where,  $\beta_s$ is the maximum surface velocity and $\xi=\displaystyle\Big(r/R\Big)$, with $r$ as the radial distance and $n$ is the flow profile, which indicates how the transverse flow velocity changes with the radial distance. Here $n$=1 corresponds to the linear flow profile. In the Tsallis blast-wave (TBW) model, the particles closer to the center of the fireball move slower than the ones at the edges. The average of the transverse velocity can be evaluated as~ \cite{PHENIX:2003wtu}, 

\begin{equation}
 <\beta> =\beta_n=\frac{\int \beta_s\xi^n\xi\;d\xi}{\int \xi\;d\xi}=\Big(\frac{2}{2+n}\Big)\beta_s.
\end{equation}
In our calculations, we have varied the parameter $n$ to explore a range of flow profiles within the Tsallis Blast-Wave model. Here, $R$ is the maximum radius of the expanding source at freeze-out ($0<\xi<1$). For the LHC energy regime, the chemical potential ($\mu$) is set to be 0 due to the the near symmetry in the particle-antiparticle production. Thus the invariant momentum spectrum for identified particles is written as,

 \begin{widetext}
\begin{equation}
\nonumber
E\frac{d^3N}{d^3\bf{p}} = \frac{d^3N}{p_Tdp_Tdyd\phi_p}= \frac{g}{\Big(2\pi\Big)^3}\int_{\sum_f} \\\Bigg[1 + (q_{AA}-1)\frac{p^\nu u_\nu}{T}\Bigg]^\frac{\displaystyle -1}{\displaystyle q_{AA}-1}p^\lambda d\sigma_\lambda, 
\end{equation}

here $\sum_f$ is the decoupling hyper-surface, $d\sigma_\lambda$ is the normal vector to the hyper-surface. Using the parametrization of the surface in the cylindrical coordinates, we write $d\sigma_\lambda$ as~\cite{Ruuskanen:1986py},

\begin{equation}
d\sigma_\lambda = (rd\phi_bdrdz, - {\bf e_r}rd\phi_bdzdt, 0, - {\bf e_z}rd\phi_bdrdt),
\end{equation}

and $p^\lambda d\sigma_\lambda$ can be expressed as,

\begin{align}
\nonumber
p^\lambda d\sigma_\lambda = rd\phi_bdy_s [m_T\tau\cosh(y_s-y)dr + m_T\sinh(y_s-y)d\tau 
- p_T\tau\cos(\phi_p)d\tau],
\end{align}
where $\displaystyle y_s = \frac{1}{2}\ln\frac{t+z}{t-z}$, $\tau = \sqrt{t^2 - z^2}$ is the longitudinal proper time. When particles
decouple at a constant time, $\tau=\tau_0$, the above equation is written as,
\begin{equation}
  p^\lambda d\sigma_\lambda = \tau_0m_T\cosh(y_s-y)rdrd\phi_bdy_s.
\end{equation}
With the expression, 
\begin{equation}
p^\nu u_\nu = m_T\cosh\rho\cosh(y_s-y) - p_T\sinh\rho\cos(\phi_p - \phi_b), 
\end{equation}
the spectrum of the identified particle can be simplified as,

\begin{align}
\nonumber
\frac{d^3N}{p_Tdp_Tdyd\phi_p} = \frac{g\tau_0}{\Big(2\pi\Big)^3}\int_{\sum_f}dy_srdrd\phi_bm_T
\times \cosh(y_s-y) \Big[1 + \frac{q_{AA}-1}{T}[m_T\cosh\rho\cosh(y_s-y) \\
- p_T\sinh\rho\cos(\phi_p - \phi_b)]\Big]^\frac{\displaystyle-1}{\displaystyle q_{AA}-1}.
\label{eq15}
\end{align}
\end{widetext}
We have used the following assumptions in the analysis of the transverse momentum spectra and azimuthal anisotropies of identified hadrons~\cite{Tang:2008ud}: 

\begin{enumerate}

\item We assume Bjorken's longitudinal expansion, which means that the measured particle yield remains independent of rapidity because we integrate over the entire source's rapidity range~\cite{PhysRevC.48.2462}. This approximately holds at mid-rapidity for RHIC and LHC energies~\cite{BRAHMS:2004adc}.

\item While we make the simplifying assumption of isotropic emission in azimuth for each local source, it's important to acknowledge that, in reality, the source's distribution may exhibit azimuthal variations or dependencies.~\cite{STAR:2000ekf}.

\item We make the assumption that the emission source maintains uniformity in both density and degree of non-equilibrium at the time of kinetic freeze-out. Nevertheless, this assumption does not hold for high-$p_T$ particles (jets) as they often demonstrate emission patterns concentrated on the surface, deviating from the assumed uniformity~\cite{Loizides:2006cs, Zhang:2007ja}.

\end{enumerate}

We have not included the contributions from the resonance decay while analysing the $p_T$- spectra of stable particle as it plays a significant role only at very low $p_T$. The detailed resonance decay kinematics and its effect on the spectra have been studied in the references~\cite{PhysRevC.48.2462} and~\cite{STAR:2008med}. Considering the above assumptions and taking $\phi_p - \phi_b$ = $\phi$  equation~\ref{eq15} becomes~\cite{Tang:2008ud},

\begin{widetext}
\begin{align}
\nonumber
\frac{d^3N}{2\pi p_Tdp_T} = \frac{g\tau_0m_T}{\Big(2\pi\Big)^3}\int_{-Y}^{+Y}\cosh(y)dy\int_{-\pi}^{+\pi}d\phi
\times \int_{0}^Rrdr\Big[1 + \frac{(q_{AA}-1)}{T}[m_T\cosh(\rho)\cosh(y) \\ 
- p_T\sinh(\rho)\cos(\phi)]\Big]^\frac{\displaystyle -1}{\displaystyle q_{AA}-1}.
\label{eq18}
\end{align}

Here, we have used Jacobian for the transformation of the coordinates and integrated it over $d\phi_p$. $Y$ is the rapidity of the emitting beam. At mid-rapidity i.e y $\simeq$ 0 above equation becomes,

\begin{align}
\label{tbw}
\nonumber
f_{eq}=\frac{d^3N}{2\pi p_Tdp_Tdy} = \frac{g\tau_0m_T}{\Big(2\pi\Big)^3}\int_{-\pi}^{+\pi}d\phi
\times \int_{0}^Rrdr\Big[1 + \frac{(q_{AA}-1)}{T}[m_T\cosh(\rho) \\
- p_T\sinh(\rho)\cos(\phi)]\Big]^\frac{\displaystyle -1}{\displaystyle q_{AA}-1}.
\end{align}
\end{widetext}

In this analysis, the initial distribution is parametrized by Tsallis distribution function~\cite{Cleymans:2011in},
\begin{equation}
f_{in}= D\left[1+(q_{pp}-1)\,\frac{m_T}{T_{ts}} \right]^{\frac{-q_{pp}}{q_{pp}-1}}.
\label{fin}
\end{equation}
Here, $\displaystyle D=\frac{gVm_{T}}{(2\pi)^2}$. V is the volume of the fireball formed in the heavy- ion collisions. Here, $T_{ts}$ is the Tsallis temperature which does not correspond to the physical temperature of individual particles but represents the non-extensive nature of the system. Its value influences the shape of the particle momentum distributions. $q_{pp}$ is another parameter in the initial Tsallis distribution of particles which indicates the degree of non-extensivity. Consequently, we have employed the Tsallis distribution to derive both the final particle distribution and the $n^{\prime th}$ anisotropic flow, $v_{n^\prime}$. This thermodynamically consistent Tsallis distribution has been utilized to investigate particle distributions arising from proton-proton collisions, as elaborated in the reference~\cite{Cleymans:2011in}. Using equations~\ref{tbw} and~\ref{fin} in equation~\ref{eq6} and taking $\tau_0 \approx V$ as a constant parameter, the final distribution can be expressed as,

\begin{widetext}
\begin{multline}
\label{ffin}
f_{fin} = D\Bigg[\frac{1}{2\pi}\int_{-\pi}^{+\pi}d\phi
\times \int_{0}^Rrdr \Big[1 + \frac{(q_{AA}-1)}{T}[m_T\cosh(\rho)- p_T\sinh(\rho)\cos(\phi)]\Big]^\frac{\displaystyle -1}{\displaystyle q_{AA}-1}\\ 
+ \Bigg(\left[1+(q_{pp}-1)\,\frac{m_T}{T_{ts}} \right]^{\frac{\displaystyle -q_{pp}}{\displaystyle q_{pp}-1}} - \Bigg(\frac{1}{2\pi}\int_{-\pi}^{+\pi}d\phi \times \int_{0}^R rdr\Big[1 + \frac{(q_{AA}-1)}{T}[m_T\cosh(\rho)\\- p_T\sinh(\rho)\cos(\phi)]\Big]^\frac{\displaystyle -1}{\displaystyle q_{AA}-1}\Bigg)\Bigg) \exp^{-t_f/\tau}\Bigg]
\end{multline}

Using equation~\ref{fin} and equation~\ref{ffin}, we calculate $v_{n^\prime}$ for the observed identified hadrons as follows:
\begin{equation}
 v_{n^\prime}(p_T)  = \frac{P}{Q},
\end{equation}
where, 

\begin{multline}
  P =  D\int d\phi\cos(n^{\prime}\phi)\Bigg[\frac{1}{2\pi}\int_{-\pi}^{+\pi}d\phi
\times \int_{0}^Rrdr\Big[1 + \frac{(q_{AA}-1)}{T}[m_T\cosh(\rho)\\- p_T\sinh(\rho)\cos(\phi)]\Big]^\frac{\displaystyle -1}{\displaystyle q_{AA}-1}
+ \Bigg(\left[1+(q_{pp}-1)\,\frac{m_T}{T_{ts}} \right]^{\frac{\displaystyle -q_{pp}}{\displaystyle q_{pp}-1}} - \Bigg(\frac{1}{2\pi}\int_{-\pi}^{+\pi}d\phi \\
\times \int_{0}^R rdr\Big[1 + \frac{(q_{AA}-1)}{T}[m_T\cosh(\rho)- p_T\sinh(\rho)\cos(\phi)]\Big]^\frac{\displaystyle -1}{\displaystyle q_{AA}-1}\Bigg)\Bigg) \exp^{-t_f/\tau}\Bigg]
\end{multline}

\begin{multline}
   Q = D\int d\phi \Bigg[\frac{1}{2\pi}\int_{-\pi}^{+\pi}d\phi
\times \int_{0}^Rrdr\Big[1 + \frac{(q_{AA}-1)}{T}[m_T\cosh(\rho)\\- p_T\sinh(\rho)\cos(\phi)]\Big]^\frac{\displaystyle -1}{\displaystyle q_{AA}-1}
+ \Bigg(\left[1+(q_{pp}-1)\,\frac{m_T}{T_{ts}} \right]^{\frac{\displaystyle -q_{pp}}{\displaystyle q_{pp}-1}} - \Bigg(\frac{1}{2\pi}\int_{-\pi}^{+\pi}d\phi \\
\times \int_{0}^R rdr\Big[1 + \frac{(q_{AA}-1)}{T}[m_T\cosh(\rho)- p_T\sinh(\rho)\cos(\phi)]\Big]^\frac{\displaystyle -1}{\displaystyle q_{AA}-1}\Bigg)\Bigg) \exp^{-t_f/\tau}\Bigg]
\end{multline}
\end{widetext}

Next, we transition to the results and discussion section to assess the efficacy of our current formalism in accurately characterizing anisotropic flow phenomena at LHC energies.

\begin{figure*}[htb]
\subfigure[\ $v_2$ for pions at Pb-Pb collisions for the 0-5$\%$ and 60-70$\%$ centralities.]{
\label{} 
\begin{minipage}[b]{0.48\textwidth}
\centering \includegraphics[width=\linewidth]{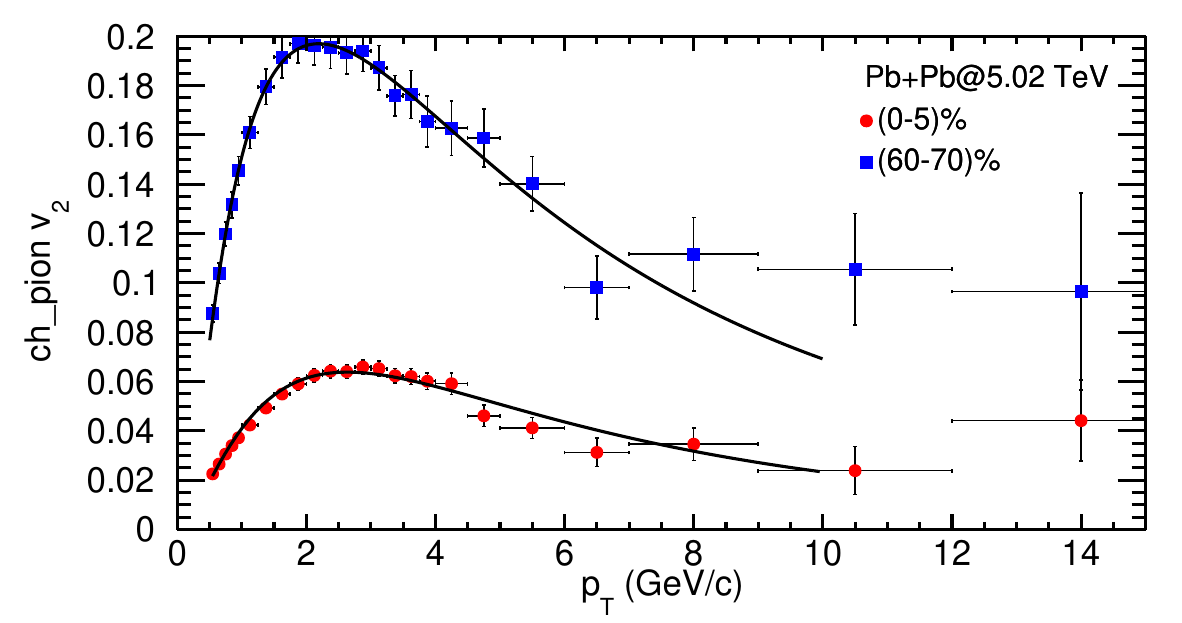}
\end{minipage}}
\hfill
\subfigure[\ $v_2$ for protons at Pb-Pb collisions for the 0-5$\%$ and 60-70$\%$ centralities.]{
\label{} 
\begin{minipage}[b]{0.48\textwidth}
\centering \includegraphics[width=\linewidth]{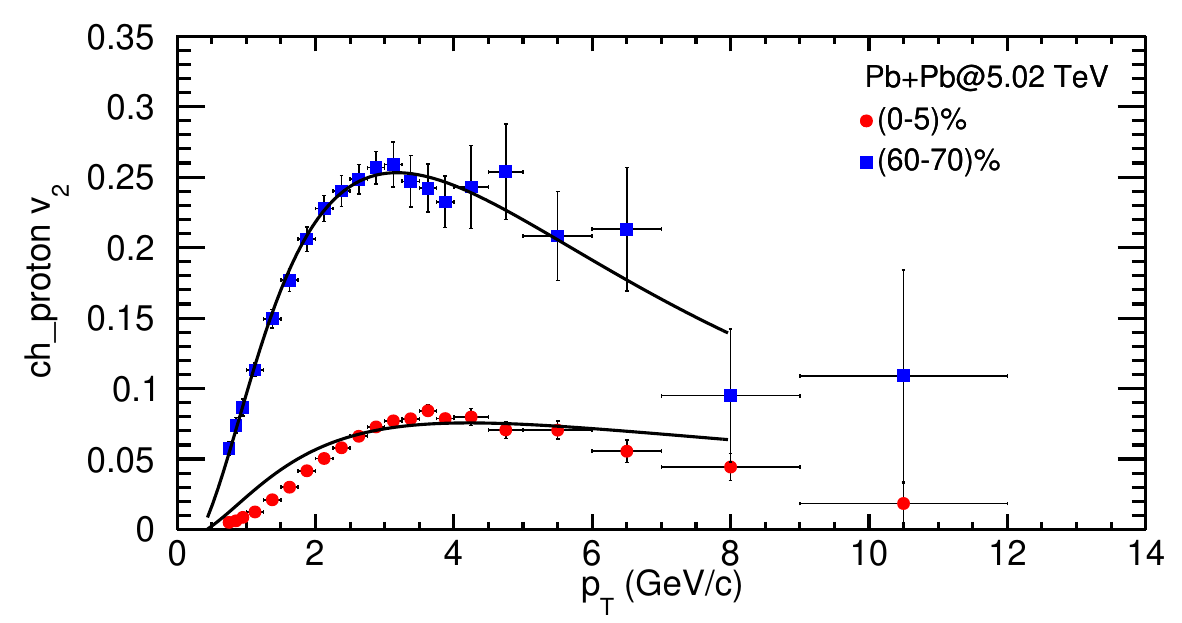}
\end{minipage}}
\hfill
\subfigure[\ $v_2$ for kaons at Pb-Pb collisions for the 0-5$\%$ and 60-70$\%$ centralities.]{
\label{fig5b} 
\begin{minipage}[b]{0.48\textwidth}
\centering \includegraphics[width=\linewidth]{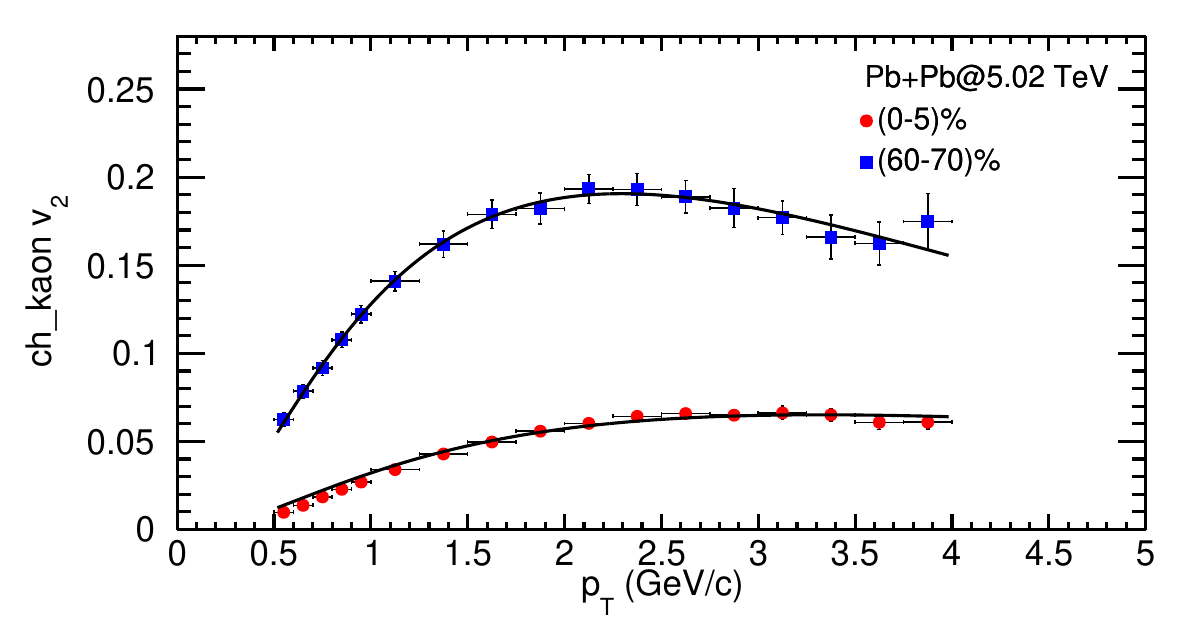}
\end{minipage}}
\hfill
\subfigure[\ $v_2$ for lambda at Pb-Pb collisions for the 0-5$\%$ and 50-60$\%$ centralities.]{
\label{fig5b} 
\begin{minipage}[b]{0.48\textwidth}
\centering \includegraphics[width=\linewidth]{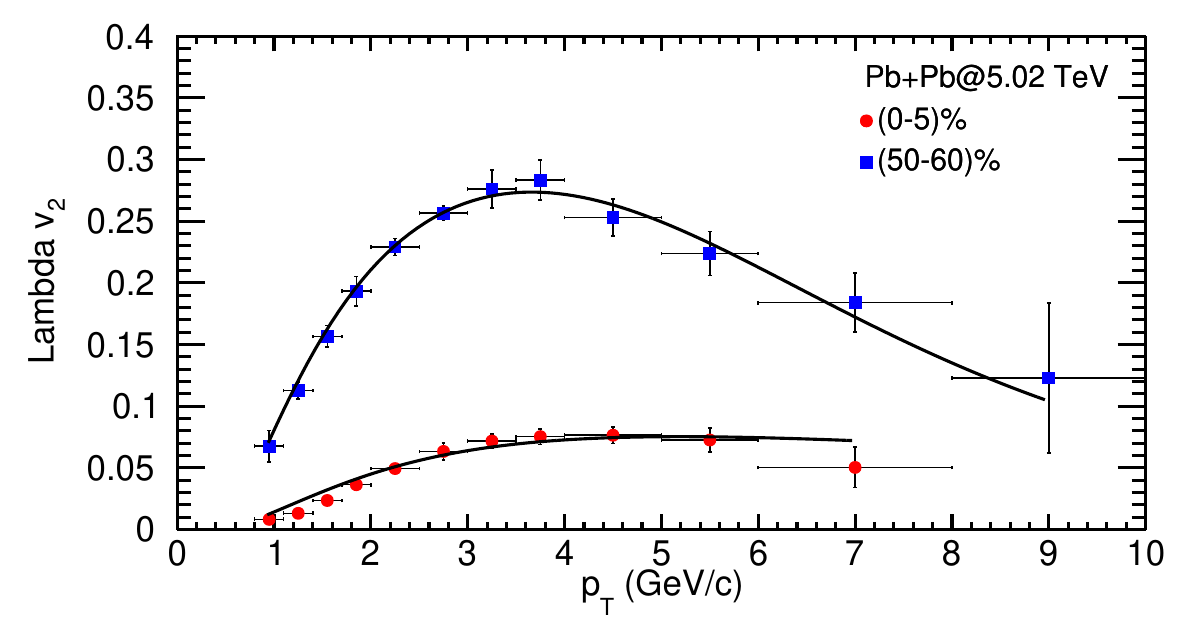}
\end{minipage}}
\hfill
\subfigure[\ $v_2$ for Omega at Pb-Pb collisions for the 10-20$\%$ and 40-50$\%$ centralities.]{
\label{fig5b} 
\begin{minipage}[b]{0.48\textwidth}
\centering \includegraphics[width=\linewidth]{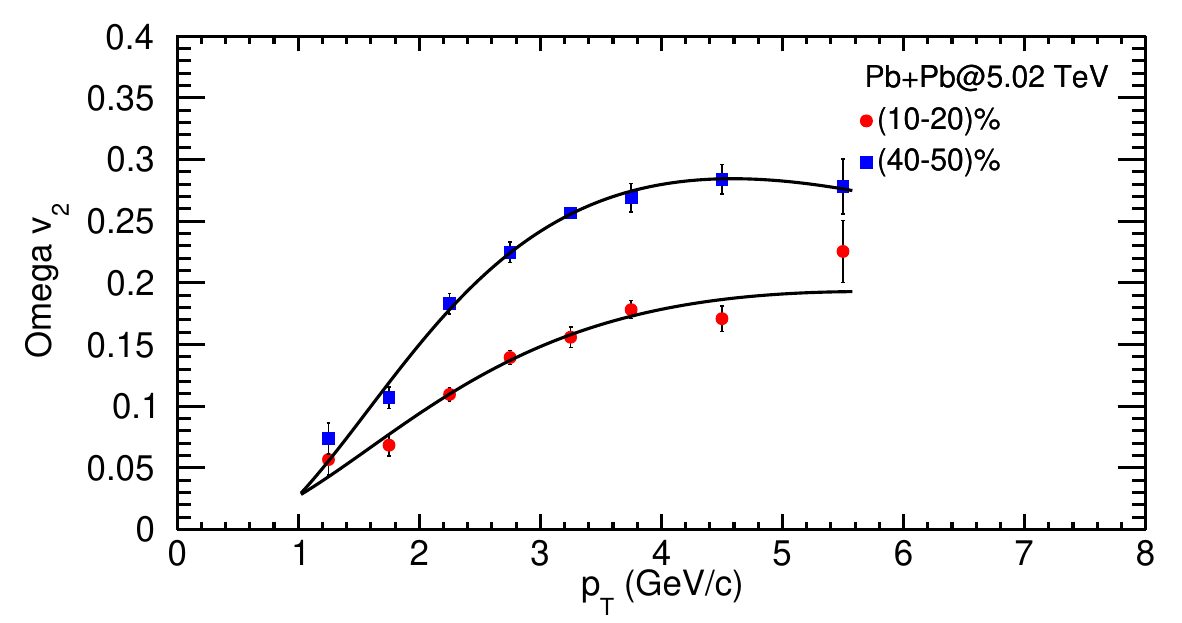}
\end{minipage}}
\caption{Elliptic flow of pions, kaons, protons, lambda and omega for Pb-Pb collisions at $\sqrt{s_{NN}}$ = 5.02 TeV.}
\label{fig3} 
\end{figure*}

\begin{figure*}[htb]
\subfigure[\ $v_2$ for pions in Xe-Xe collisions for the 0-5$\%$ and 50-60$\%$ centrality classes.]{
\label{ } 
\begin{minipage}[b]{0.48\textwidth}
\centering \includegraphics[width=\linewidth]{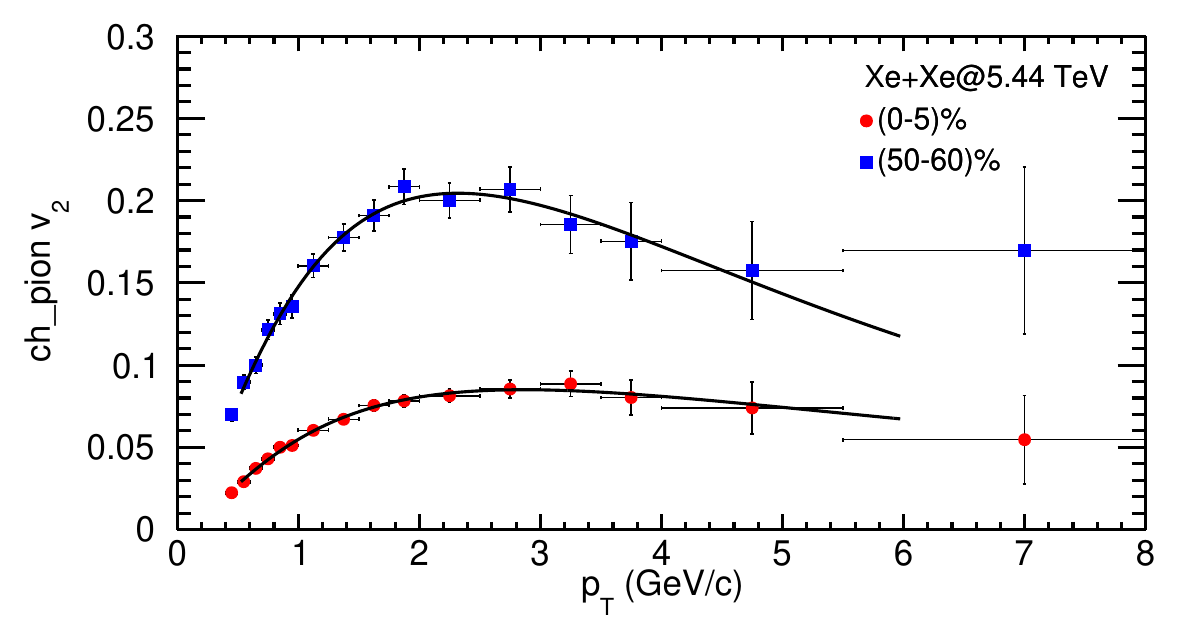}
\end{minipage}}
\hfill
\subfigure[\ $v_2$ for protons in Xe-Xe collisions for the 0-5$\%$ and 50-60$\%$ centrality classes.]{
\label{ } 
\begin{minipage}[b]{0.48\textwidth}
\centering \includegraphics[width=\linewidth]{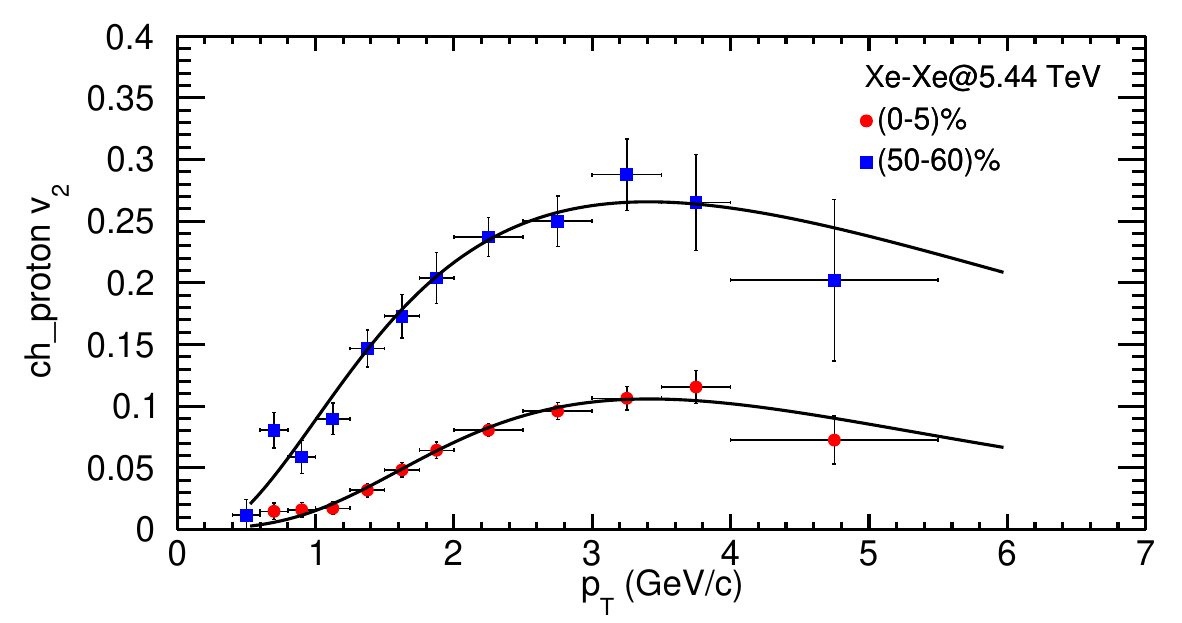}
\end{minipage}}
\hfill
\subfigure[\ $v_2$ for kaons in Xe-Xe collisions for the 0-5$\%$ and 50-60$\%$ centrality classes.]{
\label{fig5b} 
\begin{minipage}[b]{0.48\textwidth}
\centering \includegraphics[width=\linewidth]{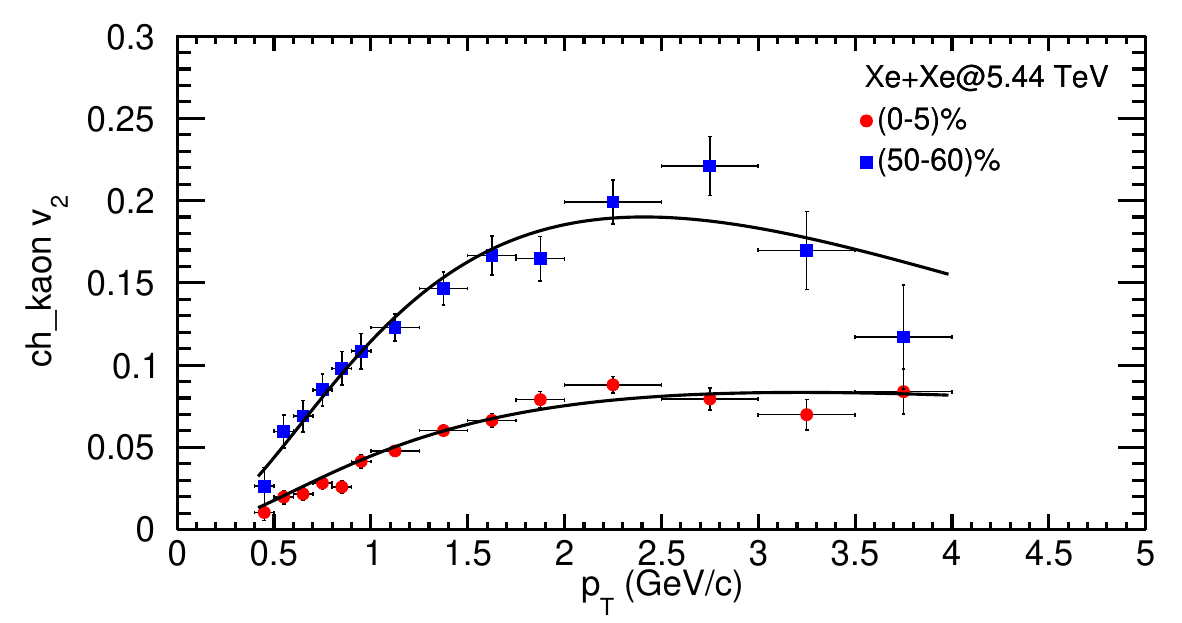}
\end{minipage}}
\caption{Elliptic flow of pions, kaons and protons for Xe-Xe collisions at $\sqrt{s_{NN}}$ = 5.44 TeV.}
\label{fig4} 
\end{figure*}

\section{Results and Discussions}
\label{RD}
Now, we proceed towards the more detailed analysis of the experimental data of transverse momentum ($p_T$) spectra, $v_2$, $v_3$ and $v_4$ measured at LHC energies for various collision systems as well as centralities. First, we analyse the experimental data of $p_T$ spectra of identified hadrons such as $\pi^{\pm}$, $K^{\pm}$ and protons for Pb-Pb and Xe-Xe collisions at $\sqrt{s_{NN}}$ = 5.02 TeV and $\sqrt{s_{NN}}$ = 5.44 TeV, respectively. We have fitted the experimental data using the equation~\ref{ffin}. Here, we consider the single freeze-out hyper-surface for all the identified hadrons. Thus, the kinetic freeze-out temperature ($T$) is considered same for all the particles and we have observed decreasing trend of $T$ when moving towards the peripheral centrality~\cite{Tang:2008ud,article}. $T$ is considered as a fixed parameter and are 0.110 GeV and 0.106 GeV for Pb-Pb collision and Xe-Xe collision, respectively for the most central case. For the peripheral collisions of Pb-Pb nuclei and Xe- Xe nuclei the observed $T$ are 0.096 GeV and 0.090 GeV, respectively. We have fitted the experimental data using the TF1 class~\cite{cernroot} available in the ROOT library~\cite{rootmanual} to get a convergent solution. The convergent solution is obtained by the $\chi^2$-minimization technique which is also used in ref.~\cite{STAR:2006vcp}.

\begin{figure*}[htb]
\subfigure[\ $v_3$ for pions in Pb-Pb collisions for the two centrality classes, 0-5$\%$ and 40-50$\%$.]{
\label{ } 
\begin{minipage}[b]{0.48\textwidth}
\centering \includegraphics[width=\linewidth]{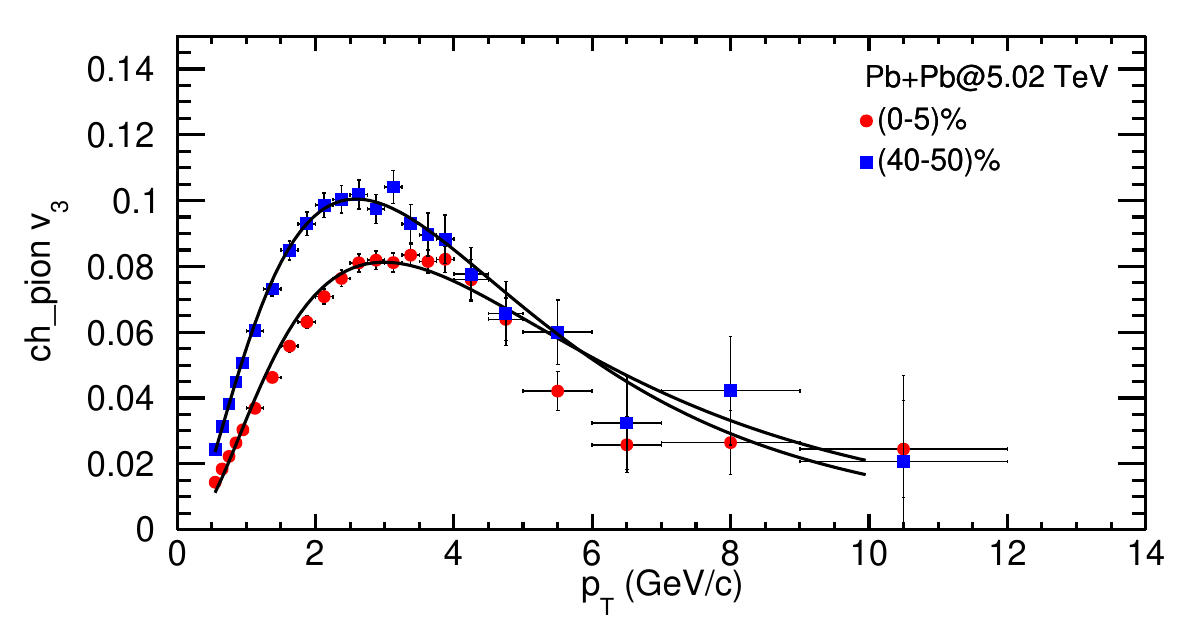}
\end{minipage}}
\hfill
\subfigure[\ $v_3$ for protons in Pb-Pb collisions for the two centrality classes, 0-5$\%$ and 40-50$\%$.]{
\label{ } 
\begin{minipage}[b]{0.48\textwidth}
\centering \includegraphics[width=\linewidth]{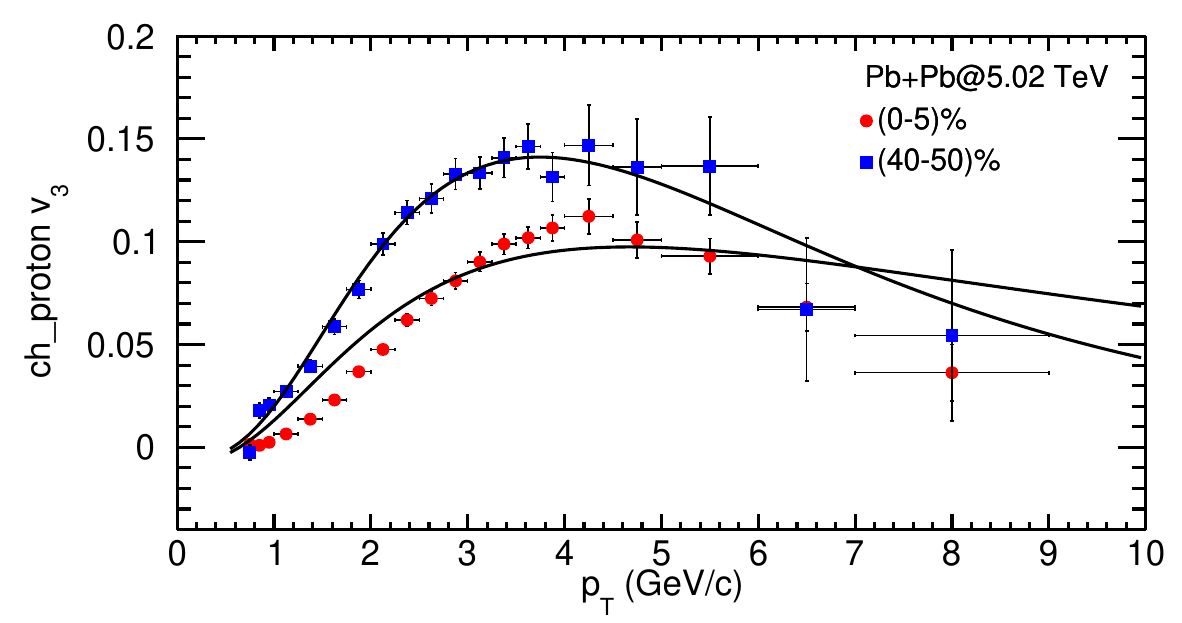}
\end{minipage}}
\hfill
\subfigure[\ $v_3$ for kaons in Pb-Pb collisions for the two centrality classes, 0-5$\%$ and 40-50$\%$.]{
\label{fig5b} 
\begin{minipage}[b]{0.48\textwidth}
\centering \includegraphics[width=\linewidth]{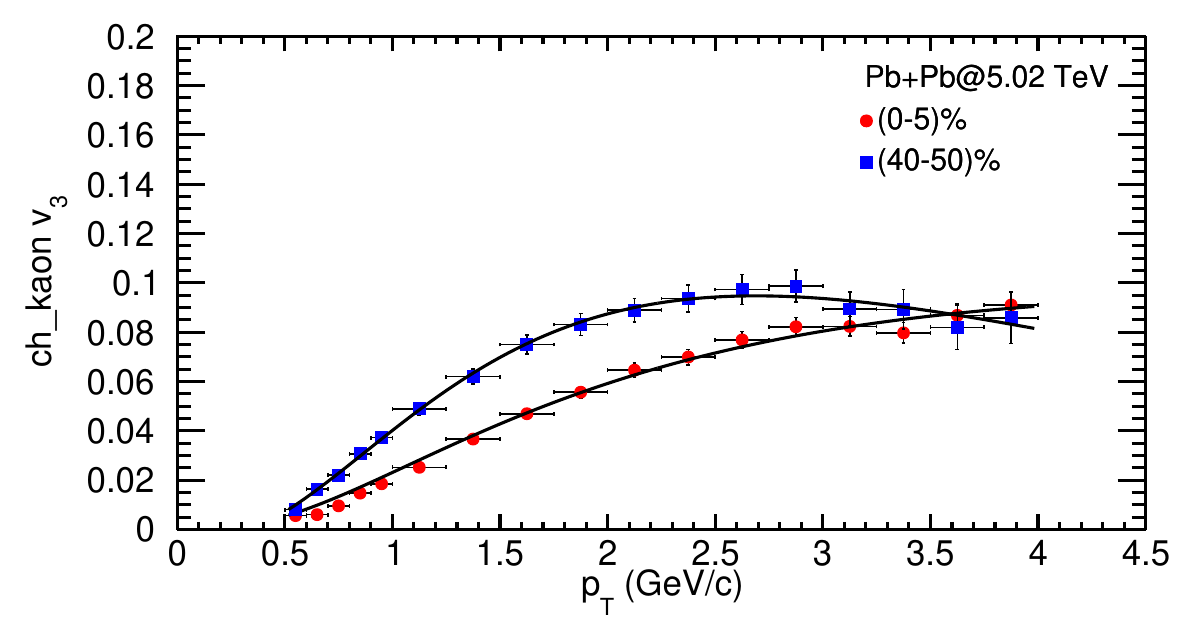}
\end{minipage}}
\hfill
\subfigure[\ $v_3$ for lambda in Pb-Pb collisions for the two centrality classes, 0-5$\%$ and 40-50$\%$.]{
\label{fig5b} 
\begin{minipage}[b]{0.48\textwidth}
\centering \includegraphics[width=\linewidth]{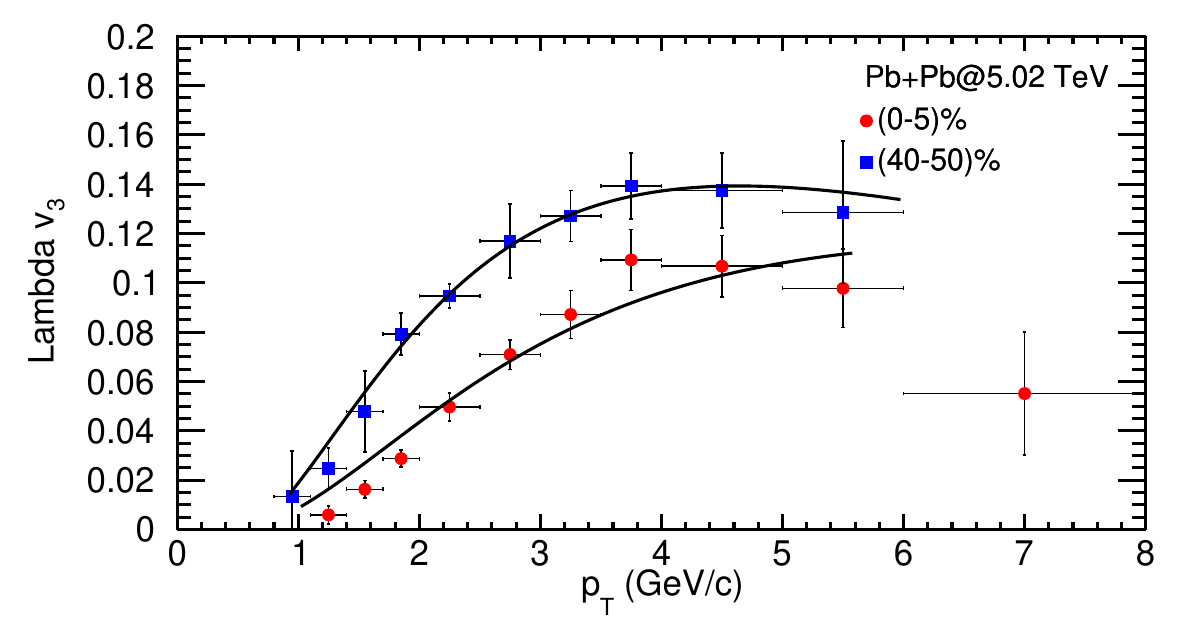}
\end{minipage}}
\caption{The triangular flow ($v_3$) of pions, kaons, protons and lambda at $\sqrt{s_{NN}}$ = 5.02 TeV for Pb- Pb collisions.}
\label{fig5} 
\end{figure*}

Figure~\ref{fig1} depicts the $p_{T}$-spectra for identified particles for the most central and peripheral Pb-Pb collisions at $\sqrt{s_{NN}}$ = 5.02 TeV~\cite{ALICE:2019hno} and Xe-Xe collisions at $\sqrt{s_{NN}}$ = 5.44 TeV~\cite{ALICE:2021lsv} fitted with our proposed formulation. Fitting of the transverse momentum spectra play a crucial role in extracting information about the particle production and dynamics in heavy-ion collisions. The choice of the fitting range is essential for obtaining meaningful results. In the low $p_T$ region, the maximum of the spectra is pushed towards higher momenta while going from peripheral to central Pb-Pb events. This effect is mass-dependent and can be interpreted as a signature of radial flow~\cite{ALICE:2013mez}. For high $p_T$, the spectra follow a power-law shape, as expected from perturbative QCD (pQCD) calculations~\cite{Kretzer:2000yf}. In our earlier work~\cite{Tripathy:2017nmo}, BTE in RTA with BGBW as equilibrium distribution function is used to fit the $p_T$- spectra and explain the data only upto $p_T$ = 5 GeV. This motivates us to use TBW as $f_{eq}$ in our present formulation in BTE with RTA. We notice that the present formulation explains the experimental data successfully upto $p_T$ = 8 GeV with a very good $\chi^2/ndf$ for all the considered identified hadrons. The value of $\chi^2/ndf$ is found to be smaller than unity because of the point-to-point systematic errors, which are included in the fit and dominate over statistical ones, are estimated on the conservative side and might not be completely random~\cite{STAR:2008med}. The extracted parameters are shown in the table~\ref{t1}. The average transverse flow velocity, $<\beta>$ decreases with the mass and also shows the decreasing trend when moving from most central towards peripheral collisions for both the Pb-Pb and Xe-Xe collisions. These findings go inline with the well established hydrodynamical behaviour. We have also noticed that, $<\beta>$ is higher for Xe-Xe collision in comparison to the Pb-Pb collisions, which suggests that the higher collision energies lead to increased particle production and multiplicity~\cite{STAR:2008med}. The collective motion and interactions of these particles can lead to a higher effective temperature within the system. We have found that the kinetic freeze-out temperature ($T$) decreases with the increasing collision energies. These findings suggest that a higher initial energy density results in a larger multiplicity and longer expansion time for the system, resulted into a large flow velocity and lower kinetic freeze-out temperature \cite{Tang:2008ud,STAR:2008med}. The kinetic freeze-out temperature ($T$) decreases from central to peripheral collisions in the present analysis which is in contrast to the findings of the conventional BGBW results~\cite{STAR:2008med}.

\begin{figure*}[htb]
\subfigure[\ $v_3$ for pions in Xe-Xe collisions for the 0-5$\%$ and 30-50$\%$ centralities.]{
\label{ } 
\begin{minipage}[b]{0.48\textwidth}
\centering \includegraphics[width=\linewidth]{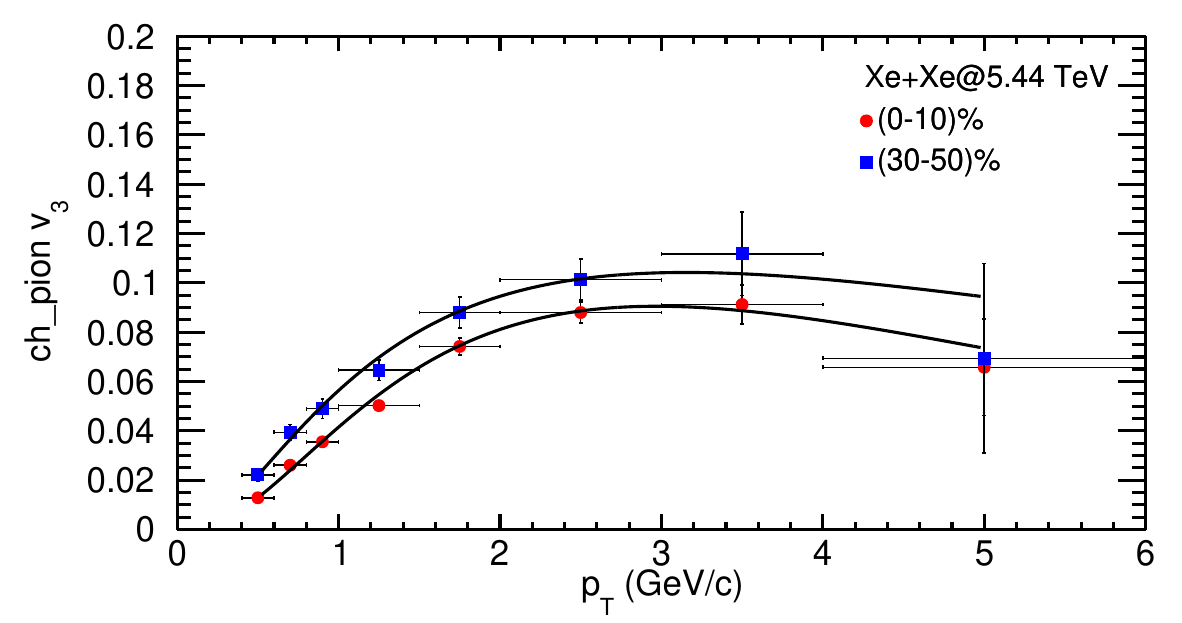}
\end{minipage}}
\hfill
\subfigure[\ $v_3$ for protons in Xe-Xe collisions for the 0-5$\%$ and 30-50$\%$ centralities.]{
\label{ } 
\begin{minipage}[b]{0.48\textwidth}
\centering \includegraphics[width=\linewidth]{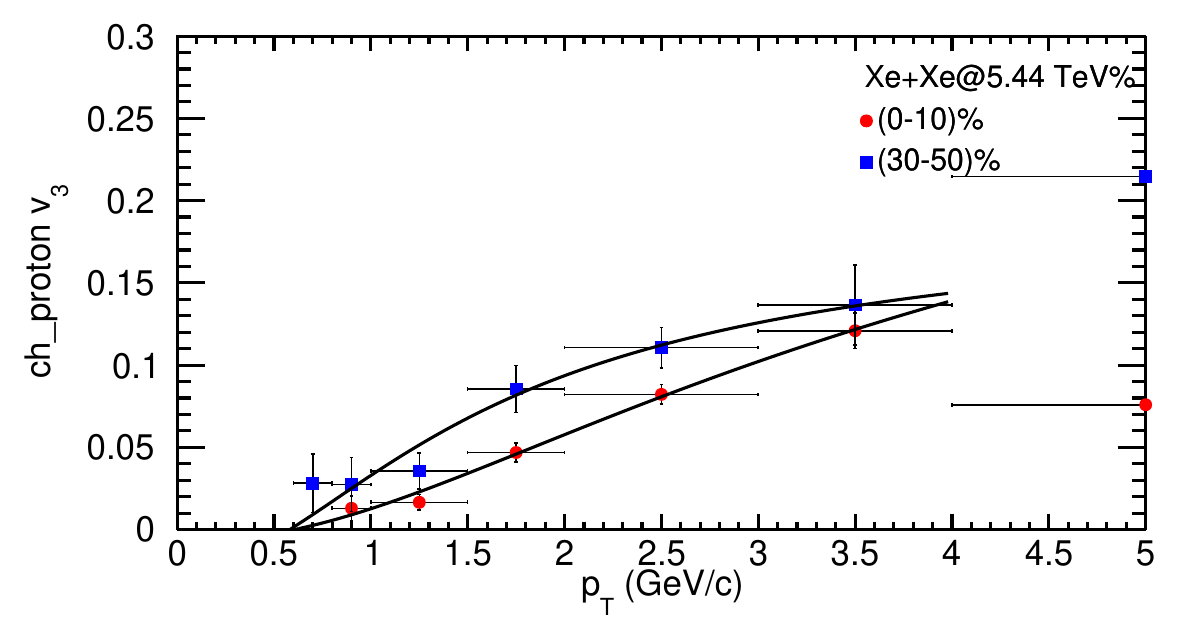}
\end{minipage}}
\hfill
\subfigure[\ $v_3$ for kaons in Xe-Xe collisions for the 0-5$\%$ and 30-50$\%$ centralities.]{
\label{fig5b} 
\begin{minipage}[b]{0.48\textwidth}
\centering \includegraphics[width=\linewidth]{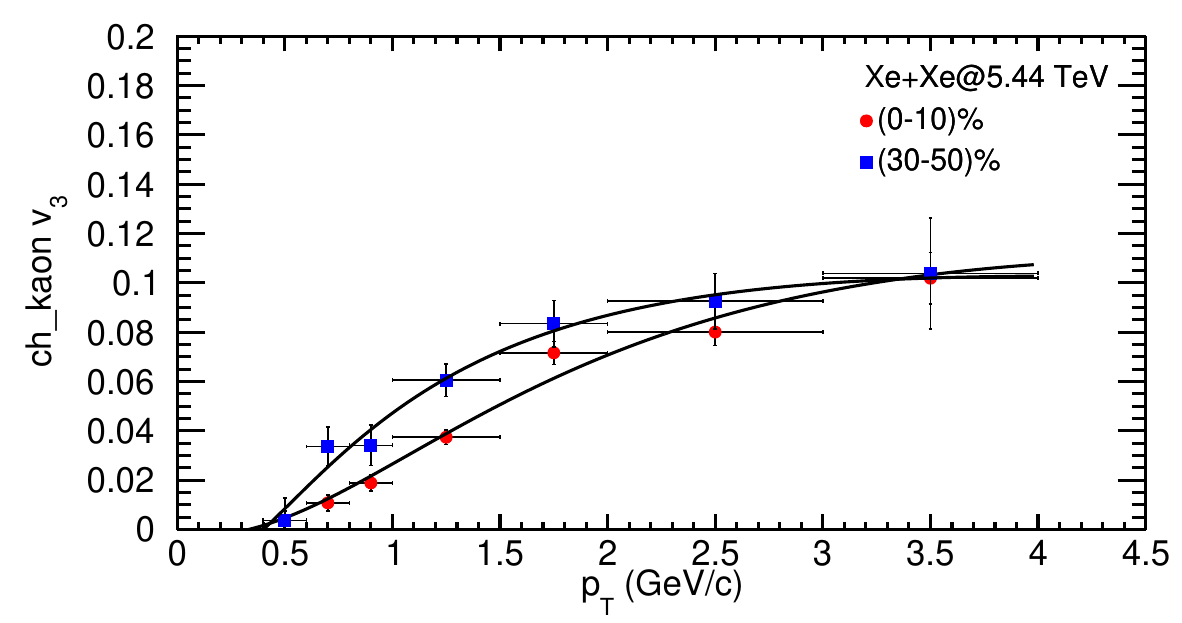}
\end{minipage}}
\caption{The triangular flow ($v_3$) of pions, kaons and protons at $\sqrt{s_{NN}}$ = 5.44 TeV for Xe-Xe collisions.}
\label{fig6} 
\end{figure*} 

The flow profile parameter $n$ increases from the most central collisions towards peripheral collisions. The large values found in peripheral collisions are maybe due to the spectrum not being thermal over the full range and increases to reproduce the power-law tail~\cite{ALICE:2013mez}. We have observed that $t_f/\tau$ increases with the mass in both the Pb-Pb and Xe-Xe collisions and does not show any centrality dependent trend which needs further investigations and will be discussed in our future work. The relationship between freeze-out time and relaxation time can vary depending on the specific details of the collision system and the assumptions made in the modelling. However, in many cases, if the freeze-out happens significantly earlier than the relaxation time, it implies that the system has not fully reached local thermal equilibrium before particles start escaping. This can happen in some high-energy or early-stage heavy ion collisions, where particles escape quickly due to the high initial energies involved. If the freeze-out time is comparable to or later than the relaxation time, it suggests that the system has had enough time to reach local thermal equilibrium before the freeze-out occurs. In this case, the observed particle spectra and properties may reflect a system that has experienced substantial equilibration and thermalization. The elevated chi-squared per degree of freedom ($\chi^2/ndf$) values suggest that the chosen range ensures a better convergence of the fitting procedure. In the high $p_T$ region ($>$ 10 GeV), the hadron production is dominated by surface emission~\cite{Zhang:2007ja} resulting in the inability of the Tsallis blast-wave model to accurately describe the spectra.

Further, we have fitted the $p_T$ spectra of pions starting from $p_T$ = 0.5 GeV as the formulation could not explain the data below this $p_T$. Pions, being among the lightest hadrons, exhibit distinct resonance effects due to their relatively small mass. We have not incorporated the contribution of pion yields from resonance decay which significantly influence the spectral shape at a very low momenta~ \cite{Che:2020fbz,PhysRevC.48.2462}. 

The anisotropic flow analysis conducted in both the Pb-Pb and Xe-Xe collisions sheds light on the intricate interplay of particle dynamics, collective effects, and collision system characteristics. By employing the Boltzmann transport equation with Tsallis distributions as the initial function and TBW as an equilibrium distribution function, a versatile framework is established to explore anisotropic flow phenomena in distinct collision systems~\cite{ALICE:2022wpn}. A pivotal accomplishment emerges as we have compared our proposed Boltzmann transport equation with the TBW as an equilibrium function to the traditional Boltzmann-Gibbs blast wave model. The former exhibits superior success in fitting anisotropic flow data as evident from the results presented in the figure~\ref{comp}, indicating its remarkable flexibility in accommodating non-equilibrium effects. Here, the fitting has been done in both the TBW and BGBW models to get the minimum value of $\chi^2/ndf$. In contrast, the limitations of Boltzmann- Gibbs blast wave model arise from its assumption of complete thermal equilibrium, potentially inhibiting its capacity to accurately represent non-equilibrium systems~\cite{PhysRevC.104.034901}.

 \begin{widetext}
\begin{figure*}[htb]
\subfigure[\ $v_4$ for pions in Pb-Pb collisions for the 0-5$\%$ and 40-50$\%$ centrality classes.]{
\label{ } 
\begin{minipage}[b]{0.48\textwidth}
\centering \includegraphics[width=\linewidth]{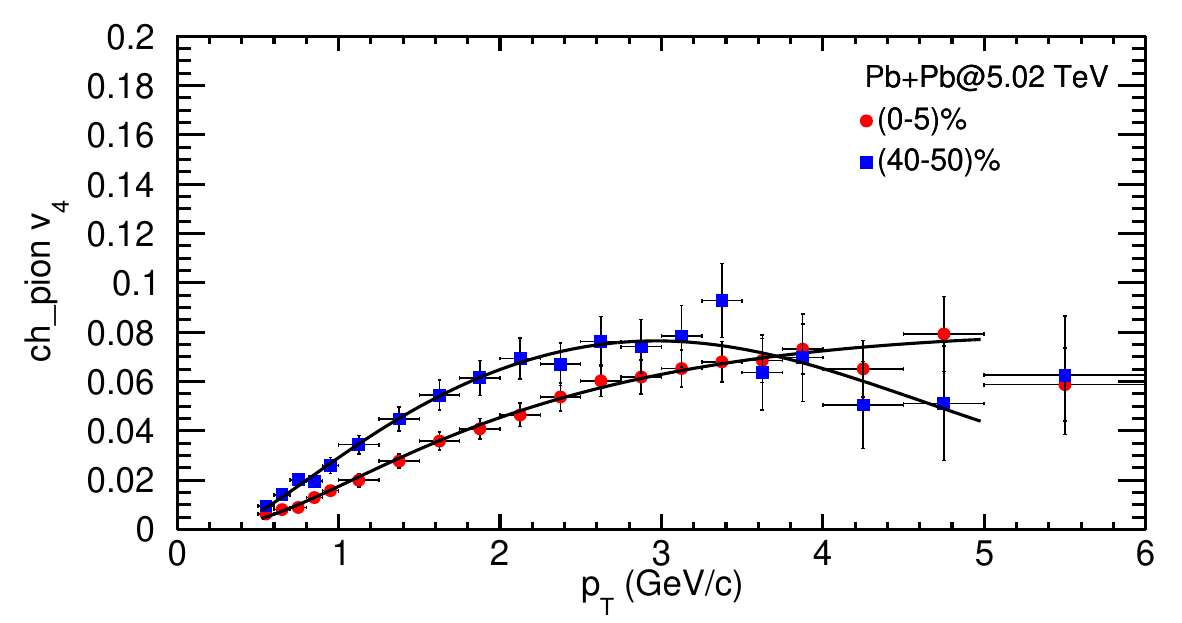}
\end{minipage}}
\hfill
\subfigure[\ $v_4$ for protons in Pb-Pb collisions for the 0-5$\%$ and 40-50$\%$ centrality classes.]{
\label{ } 
\begin{minipage}[b]{0.48\textwidth}
\centering \includegraphics[width=\linewidth]{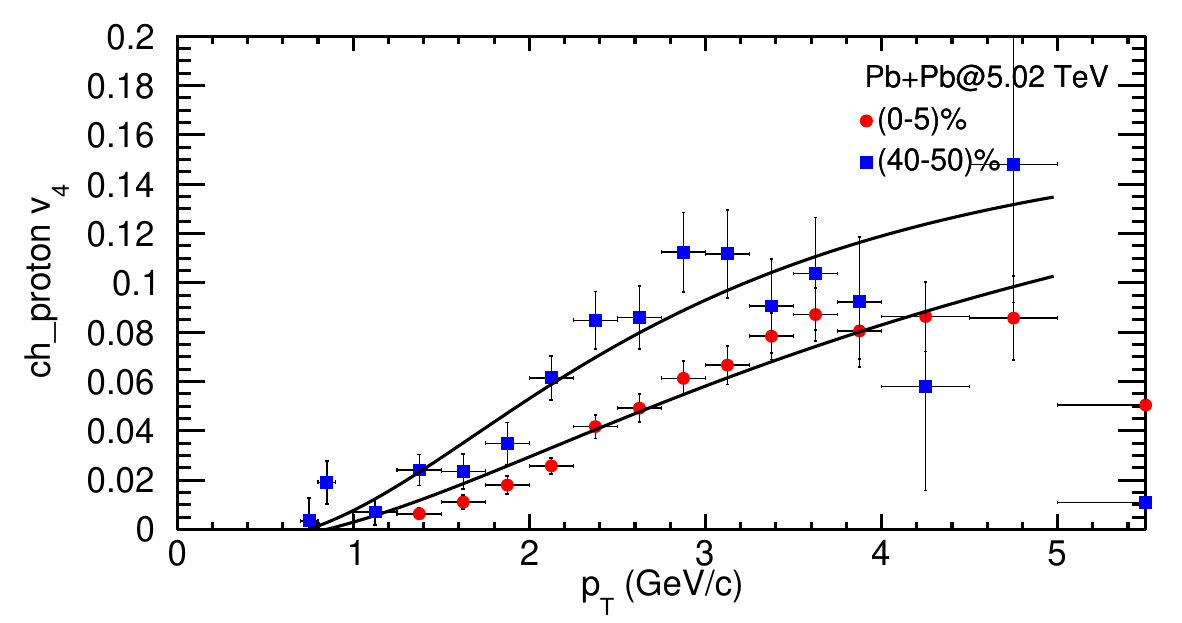}
\end{minipage}}
\hfill
\subfigure[\ $v_4$ for kaons in Pb-Pb collisions for the 0-5$\%$ and 40-50$\%$ centrality classes.]{
\label{fig5b} 
\begin{minipage}[b]{0.48\textwidth}
\centering \includegraphics[width=\linewidth]{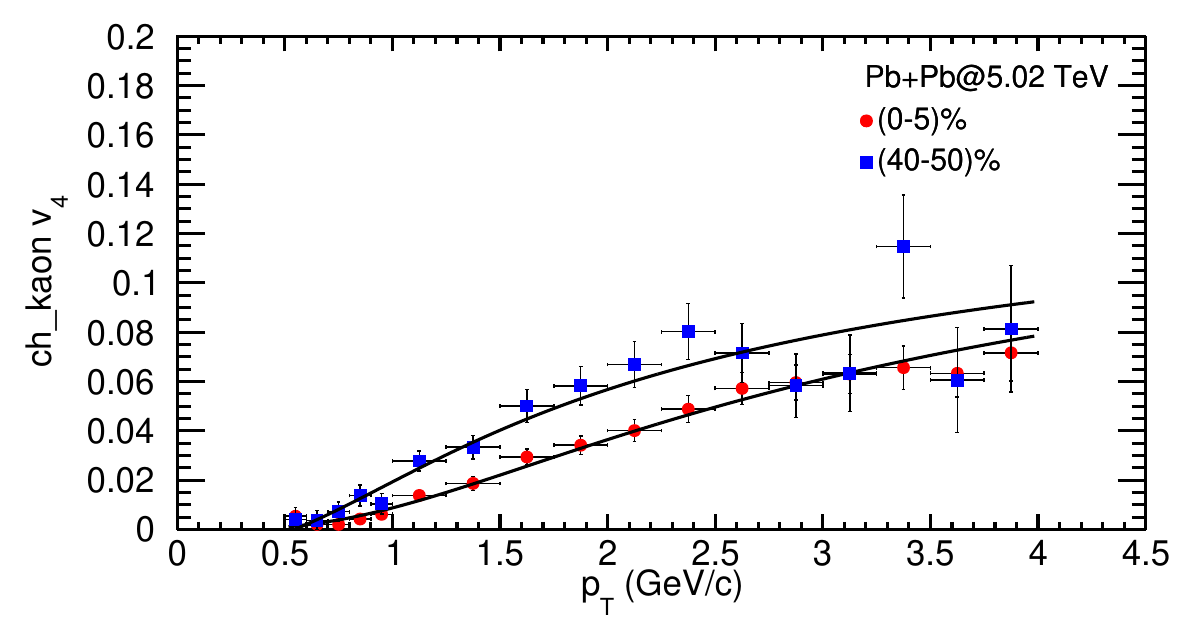}
\end{minipage}}
\caption{The quadrangular flow ($v_4$) of pions, kaons and protons at $\sqrt{s_{NN}}$ = 5.02 TeV for Pb- Pb collisions.}
\label{fig7} 
\end{figure*}

\begin{table}[]
\caption{The parameters obtained by fitting the $p_T$- spectra for Pb-Pb and Xe-Xe collisions at LHC energies for the most central and peripheral collisions.}
\label{t1}
\begin{tabular}{l|llllll|llllll|}
\cline{2-13}
\multirow{3}{*}{} &
  \multicolumn{6}{l|}{\textbf{Pb-Pb}} &
  \multicolumn{6}{l|}{\textbf{Xe-Xe}} \\ \cline{2-13} 
 &
  \multicolumn{3}{l|}{\textbf{(0-5)\%}} &
  \multicolumn{3}{l|}{\textbf{(70-80)\%}} &
  \multicolumn{3}{l|}{\textbf{(0-5)\%}} &
  \multicolumn{3}{l|}{\textbf{(60-70)\%}} \\ \cline{2-13} 
 &
  \multicolumn{1}{l|}{\textbf{${\pi^{+}+\pi^{-}}$}} &
  \multicolumn{1}{l|}{\textbf{${K^{+}+K^{-}}$}} &
  \multicolumn{1}{l|}{\textbf{${p+\bar{p}}$}} &
  \multicolumn{1}{l|}{\textbf{${\pi^{+}+\pi^{-}}$}} &
  \multicolumn{1}{l|}{\textbf{$K^{+}+K^{-}$}} &
  \textbf{$p+\bar{p}$} &
  \multicolumn{1}{l|}{\textbf{${\pi^{+}+\pi^{-}}$}} &
  \multicolumn{1}{l|}{\textbf{$K^{+}+K^{-}$}} &
  \multicolumn{1}{l|}{\textbf{$p+\bar{p}$}} &
  \multicolumn{1}{l|}{\textbf{${\pi^{+}+\pi^{-}}$}} &
  \multicolumn{1}{l|}{\textbf{$K^{+}+K^{-}$}} &
  \textbf{$p+\bar{p}$} \\ \hline
\multicolumn{1}{|l|}{\textbf{$<\beta>$}} &
  \multicolumn{1}{l|}{\textbf{0.609}} &
  \multicolumn{1}{l|}{\textbf{0.573}} &
  \multicolumn{1}{l|}{\textbf{0.565}} &
  \multicolumn{1}{l|}{\textbf{0.462}} &
  \multicolumn{1}{l|}{\textbf{0.424}} &
  \textbf{0.349} &
  \multicolumn{1}{l|}{\textbf{0.642}} &
  \multicolumn{1}{l|}{\textbf{0.632}} &
  \multicolumn{1}{l|}{\textbf{0.590}} &
  \multicolumn{1}{l|}{\textbf{0.508}} &
  \multicolumn{1}{l|}{\textbf{0.498}} &
  \textbf{0.345} \\ \hline
\multicolumn{1}{|l|}{\textbf{n}} &
  \multicolumn{1}{l|}{\textbf{0.591}} &
  \multicolumn{1}{l|}{\textbf{0.708}} &
  \multicolumn{1}{l|}{\textbf{0.453}} &
  \multicolumn{1}{l|}{\textbf{2.156}} &
  \multicolumn{1}{l|}{\textbf{1.651}} &
  \textbf{1.335} &
  \multicolumn{1}{l|}{\textbf{0.965}} &
  \multicolumn{1}{l|}{\textbf{0.794}} &
  \multicolumn{1}{l|}{\textbf{3}} &
  \multicolumn{1}{l|}{\textbf{1.83}} &
  \multicolumn{1}{l|}{\textbf{1.457}} &
  \textbf{0.39} \\ \hline
\multicolumn{1}{|l|}{\textbf{$q_{pp}$}} &
  \multicolumn{1}{l|}{\textbf{1.125}} &
  \multicolumn{1}{l|}{\textbf{1.157}} &
  \multicolumn{1}{l|}{\textbf{1.120}} &
  \multicolumn{1}{l|}{\textbf{1.165}} &
  \multicolumn{1}{l|}{\textbf{1.170}} &
  \textbf{1.139} &
  \multicolumn{1}{l|}{\textbf{1.124}} &
  \multicolumn{1}{l|}{\textbf{1.135}} &
  \multicolumn{1}{l|}{\textbf{1.141}} &
  \multicolumn{1}{l|}{\textbf{1.15}} &
  \multicolumn{1}{l|}{\textbf{1.14}} &
  \textbf{1.15} \\ \hline
\multicolumn{1}{|l|}{\textbf{$T_{ts}$}} &
  \multicolumn{1}{l|}{\textbf{0.130}} &
  \multicolumn{1}{l|}{\textbf{0.089}} &
  \multicolumn{1}{l|}{\textbf{0.120}} &
  \multicolumn{1}{l|}{\textbf{0.0872}} &
  \multicolumn{1}{l|}{\textbf{0.0876}} &
  \textbf{0.0699} &
  \multicolumn{1}{l|}{\textbf{0.130}} &
  \multicolumn{1}{l|}{\textbf{0.117}} &
  \multicolumn{1}{l|}{\textbf{0.120}} &
  \multicolumn{1}{l|}{\textbf{0.0865}} &
  \multicolumn{1}{l|}{\textbf{0.080}} &
  \textbf{0.0754} \\ \hline
\multicolumn{1}{|l|}{\textbf{$q_{AA}$}} &
  \multicolumn{1}{l|}{\textbf{1.0193}} &
  \multicolumn{1}{l|}{\textbf{1.045}} &
  \multicolumn{1}{l|}{\textbf{1.053}} &
  \multicolumn{1}{l|}{\textbf{1.10}} &
  \multicolumn{1}{l|}{\textbf{1.10}} &
  \textbf{1.094} &
  \multicolumn{1}{l|}{\textbf{1.042}} &
  \multicolumn{1}{l|}{\textbf{1.046}} &
  \multicolumn{1}{l|}{\textbf{1.0235}} &
  \multicolumn{1}{l|}{\textbf{1.10}} &
  \multicolumn{1}{l|}{\textbf{1.096}} &
  \textbf{1.08} \\ \hline
\multicolumn{1}{|l|}{\textbf{$t_f/\tau$}} &
  \multicolumn{1}{l|}{\textbf{1.1}} &
  \multicolumn{1}{l|}{\textbf{1.196}} &
  \multicolumn{1}{l|}{\textbf{2.434}} &
  \multicolumn{1}{l|}{\textbf{1.1}} &
  \multicolumn{1}{l|}{\textbf{1.1}} &
  \textbf{5} &
  \multicolumn{1}{l|}{\textbf{1.001}} &
  \multicolumn{1}{l|}{\textbf{1.521}} &
  \multicolumn{1}{l|}{\textbf{3.176}} &
  \multicolumn{1}{l|}{\textbf{1}} &
  \multicolumn{1}{l|}{\textbf{1}} &
  \textbf{3.745} \\ \hline
\multicolumn{1}{|l|}{\textbf{$\chi^2/ndf$}} &
  \multicolumn{1}{l|}{\textbf{1.145}} &
  \multicolumn{1}{l|}{\textbf{0.106}} &
  \multicolumn{1}{l|}{\textbf{1.154}} &
  \multicolumn{1}{l|}{\textbf{0.036}} &
  \multicolumn{1}{l|}{\textbf{0.0851}} &
  \textbf{0.196} &
  \multicolumn{1}{l|}{\textbf{0.264}} &
  \multicolumn{1}{l|}{\textbf{0.106}} &
  \multicolumn{1}{l|}{\textbf{0.259}} &
  \multicolumn{1}{l|}{\textbf{0.126}} &
  \multicolumn{1}{l|}{\textbf{0.215}} &
  \textbf{0.468} \\ \hline
\end{tabular}
\end{table}
\end{widetext}

In figures~\ref{fig3} and ~\ref{fig4}, we have presented the fitting of the experimental data for elliptic flow ($v_2$) of $\pi^{\pm}$, $K^{\pm}$, $p+\bar{p}$,  $\Lambda+\bar{\Lambda}$ and $\Omega^{\mp}$ at $\sqrt{s_{NN}}$ = 5.02 TeV for Pb-Pb~\cite{ALICE:2018yph,ALICE:2022zks} and  at $\sqrt{s_{NN}}$ = 5.44 TeV for Xe-Xe collisions~\cite{ALICE:2021ibz}, respectively for the most central as well as peripheral collisions. The present formulation explains the experimental data upto $p_T$ = 10 GeV for pions and upto $p_T$ = 8 GeV for protons. Our formulation fits the given experimental data for lambda successfully. For kaons, the fitting is upto only $p_T$ = 4 GeV and for omega it is upto $p_T$ = 5.6 GeV due to the unavailability of experimental data at higher transverse momentum. In Xe-Xe collisions, we have taken the experimental data upto $p_T$ = 6 GeV for pions and protons, as values beyond this threshold manifest considerable error bars. For kaons, the range remains at 4 GeV due to data point limitations. In order to emphasize the effect of the azimuthal modulation amplitude on the azimuthal flows, here we have considered the $<\beta>$ and $T$ as fixed parameters extracted from the fitting of the $p_T$- spectra. We notice that, in both the Pb-Pb as well as Xe-Xe collisions, $\rho_a$ increases as we go from most central to peripheral collisions.

\begin{figure*}[htb]
\subfigure[\ $\rho_a$ of identified hadrons extracted from the fitting of $v_2$ at $\sqrt{s_{NN}} $= 5.02 TeV for 0-5$\%$ and 60-70$\%$ centralities.]{
\label{ } 
\begin{minipage}[b]{0.48\textwidth}
\centering \includegraphics[width=\linewidth]{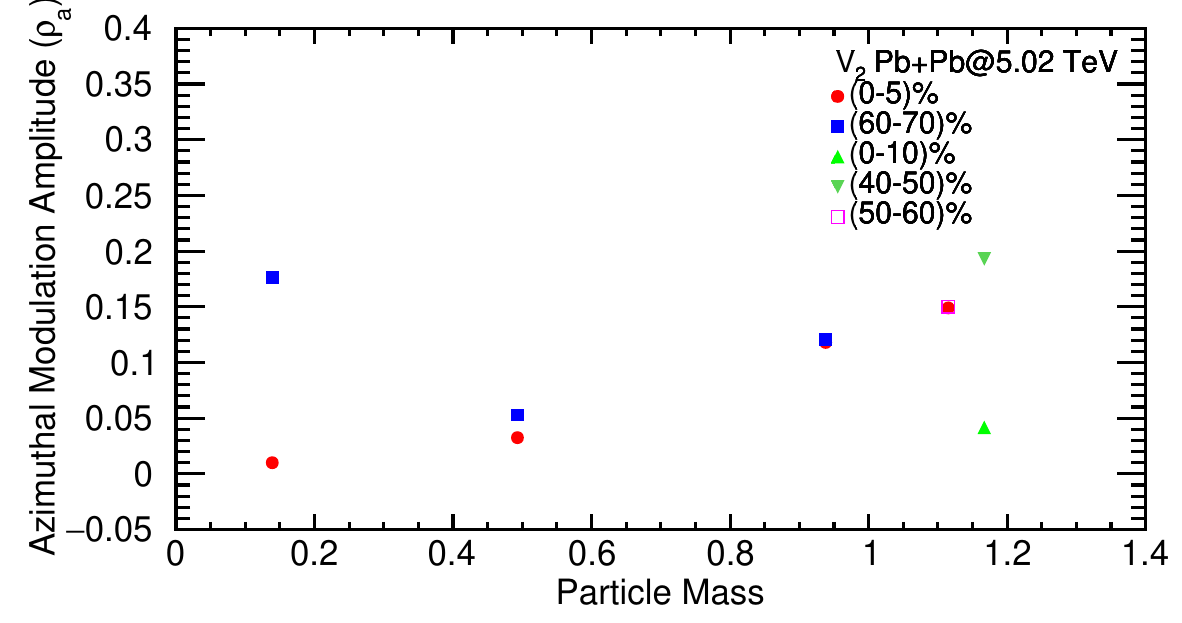}
\end{minipage}}
\hfill
\subfigure[\ $\rho_a$ of identified hadrons extracted from the fitting of $v_3$ at $\sqrt{s_{NN}} $= 5.02 TeV for 0-5$\%$ and 40-50$\%$ centralities.]{
\label{ } 
\begin{minipage}[b]{0.48\textwidth}
\centering \includegraphics[width=\linewidth]{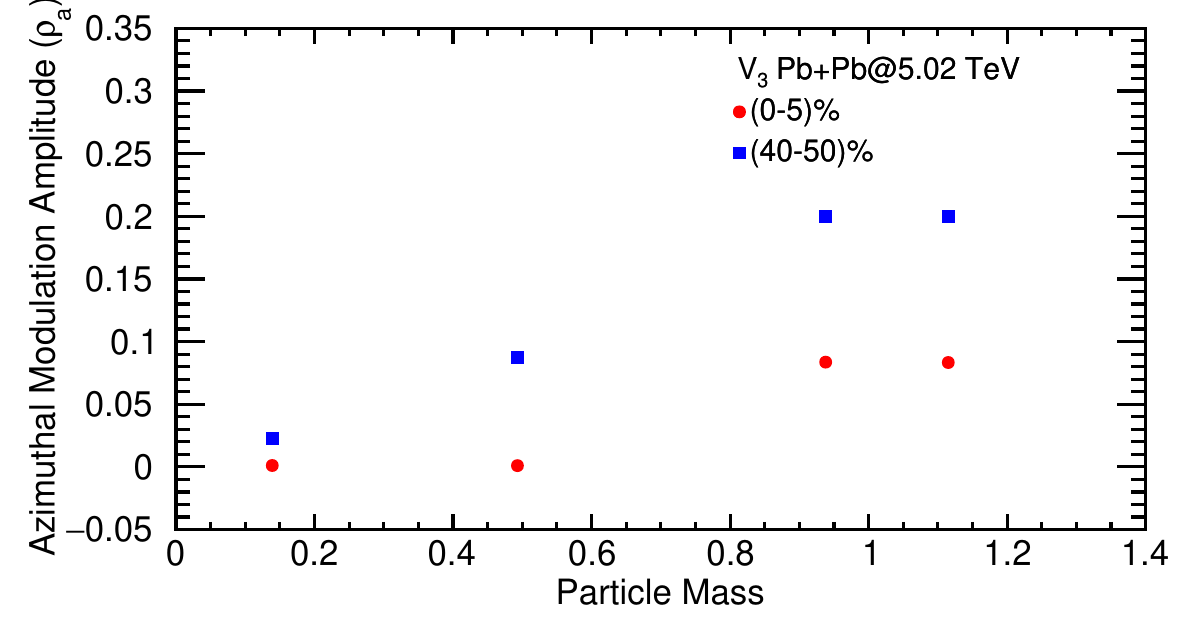}
\end{minipage}}
\hfill
\subfigure[\ $\rho_a$ of identified hadrons extracted from the fitting of $v_4$ at $\sqrt{s_{NN}} $= 5.02 TeV for 0-5$\%$ and 40-50$\%$ centralities.]{
\label{fig5b} 
\begin{minipage}[b]{0.48\textwidth}
\centering \includegraphics[width=\linewidth]{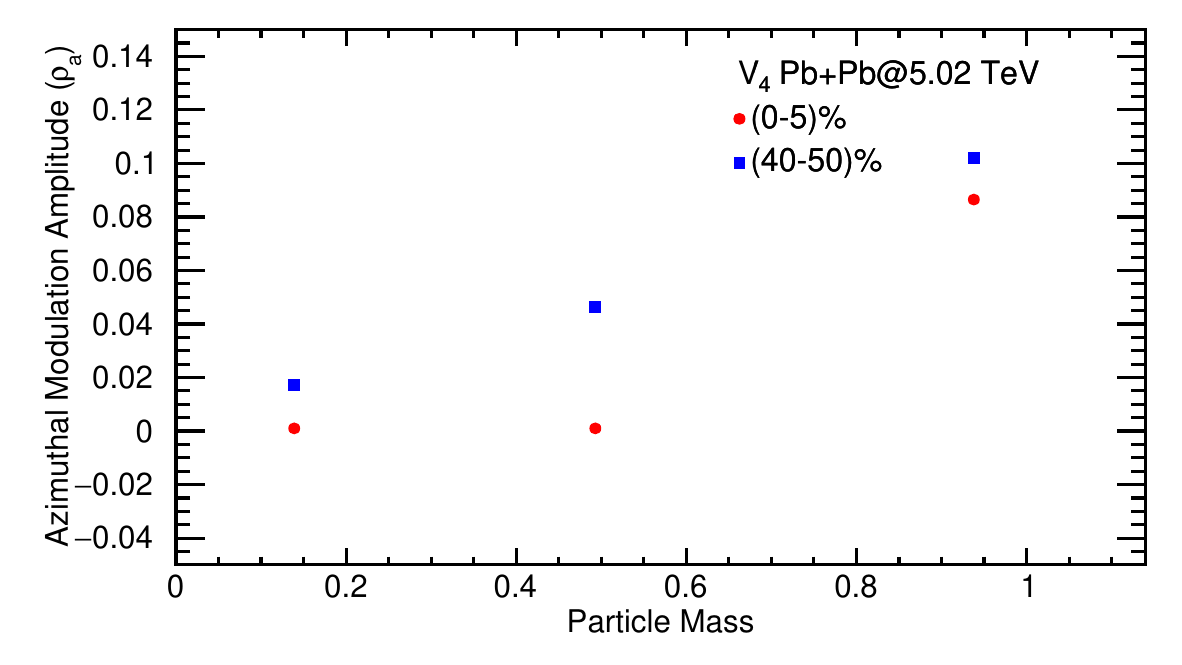}
\end{minipage}}
\caption{Variation of azimuthal modulation amplitude $\rho_a$ with the mass of the particles in Pb-Pb colisions at $\sqrt{s_{NN}} $= 5.02 TeV for the most central and peripheral collisions.}
\label{fig8} 
\end{figure*}

Figures~\ref{fig5} and ~\ref{fig6} illustrate the fitting of the triangular flow ($v_3$) of $\pi^{\pm}$, $K^{\pm}$, $p+\bar{p}$,$\Lambda+\bar{\Lambda}$ for Pb-Pb collisions at $\sqrt{s_{NN}}$ = 5.02 TeV~\cite{ALICE:2018yph} and that of $\pi^{\pm}$, $K^{\pm}$, and $p+\bar{p}$ for Xe- Xe collisions at $\sqrt{s_{NN}}$ = 5.44 TeV~\cite{ALICE:2021ibz}, respectively. In Pb-Pb collisions, we have fitted the experimental results upto $p_T$ = 10 GeV for both pions and protons. For $p_T >$ 10 GeV, we get a higher value of $\chi^2/ndf$. For kaons, the experimental data is available upto only 4 GeV so we have not fitted beyond this due to unavailability of data. We have fitted the experimental data for lambda upto  $p_T$ = 5.6 GeV as we are getting large $\chi^2/ndf$ values. We observe that, in Pb-Pb collisions, $\rho_a$ increases when one go towards the peripheral collision. In Xe-Xe collisions, we have fitted the experimental results upto $p_T$ = 5 GeV for pions while for kaons it is upto $p_T$ = 4 GeV imposed by the limitations of data points within the given dataset. For protons, we have considered the experimental data upto $p_T$ = 4 GeV as beyond this we get higher $\chi^2/ndf$ values. We again notice that $\rho_a$ increases with the centrality for all the particles. 

In figure~\ref{fig7}, we have displayed the fitting of the experimental data for $v_4$ as a function of $p_T$ for Pb-Pb collisions at $\sqrt{s_{NN}}$ = 5.02 TeV~\cite{ALICE:2018yph} for the various centralities with a $p_T$ range upto 5 GeV for pions and protons. This limitation is attributed to the presence of large error bars in the experimental data, which resulted in a large value of $\chi^2/ndf$. For kaons, the range is also limited to 4 GeV, in accordance with the available data. Again, We find that $\rho_a$ increases when one goes from the most central to peripheral collisions.

\begin{figure*}[htb]
\subfigure[\ $\rho_a$ of identified hadrons extracted from the fitting of $v_2$ at $\sqrt{s_{NN}} $= 5.44 TeV for 0-5$\%$ and 50-60$\%$ centrality classes.]{
\label{ } 
\begin{minipage}[b]{0.48\textwidth}
\centering \includegraphics[width=\linewidth]{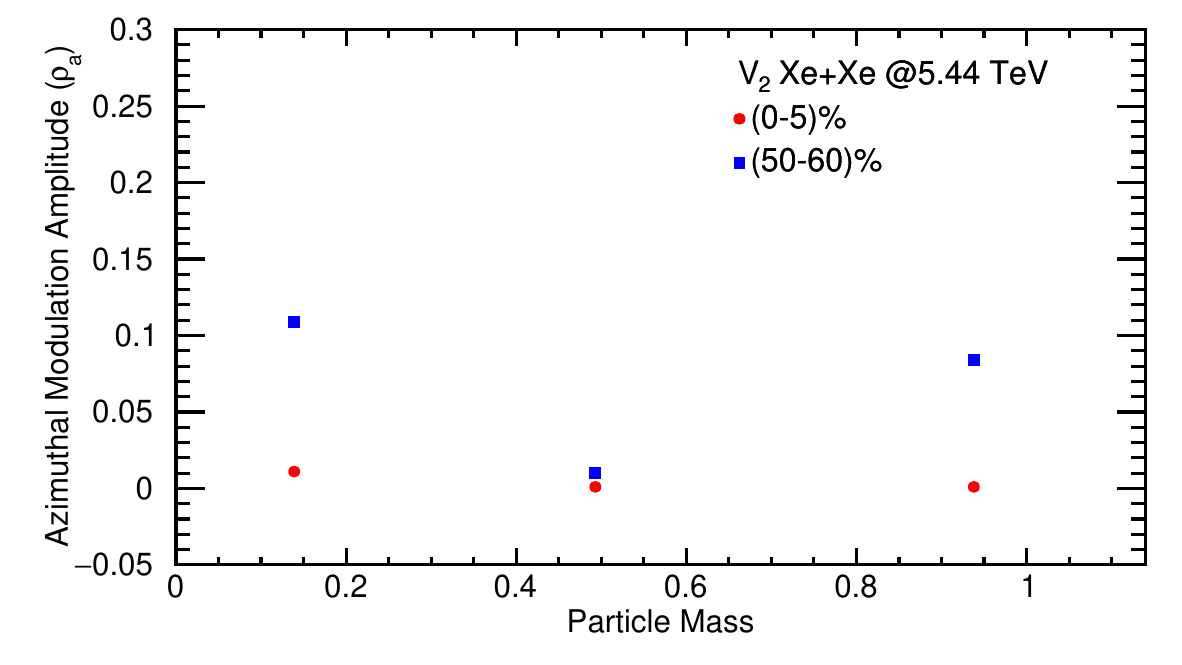}
\end{minipage}}
\hfill
\subfigure[\ $\rho_a$ of identified hadrons extracted from the fitting of $v_3$ at $\sqrt{s_{NN}} $= 5.44 TeV for 0-5$\%$ and 30-50$\%$ centrality classes.]{
\label{ } 
\begin{minipage}[b]{0.48\textwidth}
\centering \includegraphics[width=\linewidth]{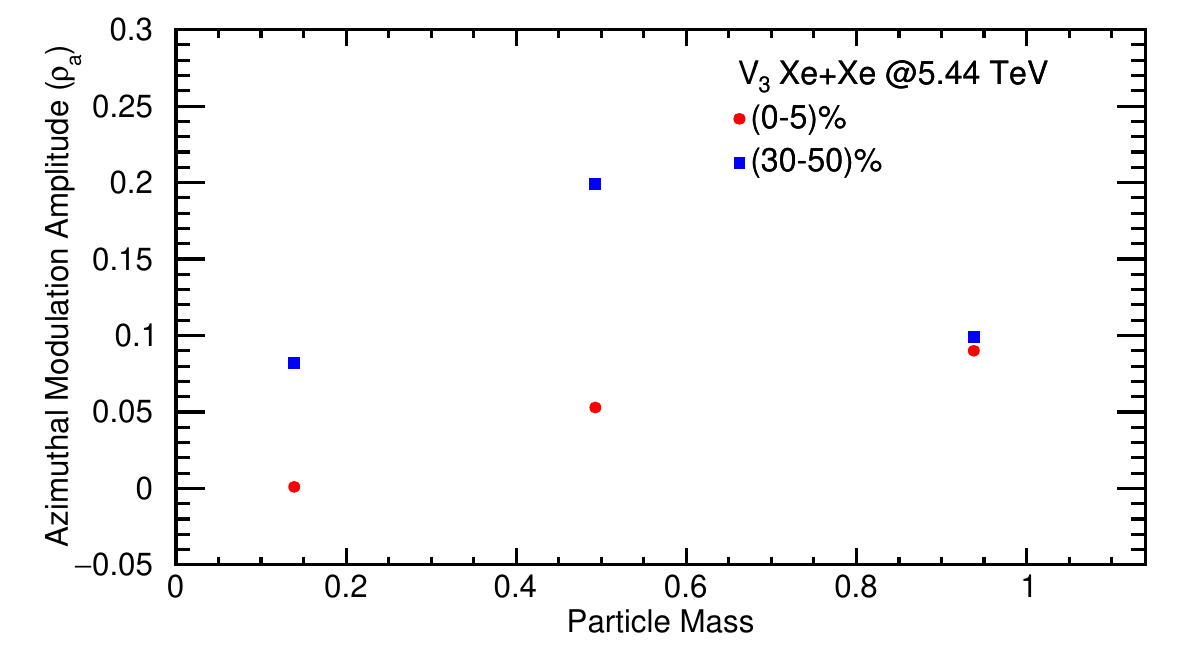}
\end{minipage}}
\caption{Variation of azimuthal modulation amplitude $\rho_a$ with the mass of the particles in Xe-Xe colisions at $\sqrt{s_{NN}} $= 5.44 TeV for the most central and peripheral collisions.}
\label{fig9} 
\end{figure*}

\begin{figure*}[htb]
\subfigure[\ $v_2$ of $\pi^{\pm}$ as a function of $p_T$ compared with the hydrodynamical models at $\sqrt{s_{NN}} $= 5.02 TeV for 0-5$\%$ centrality.]{
\label{ } 
\begin{minipage}[b]{0.48\textwidth}
\centering \includegraphics[width=\linewidth]{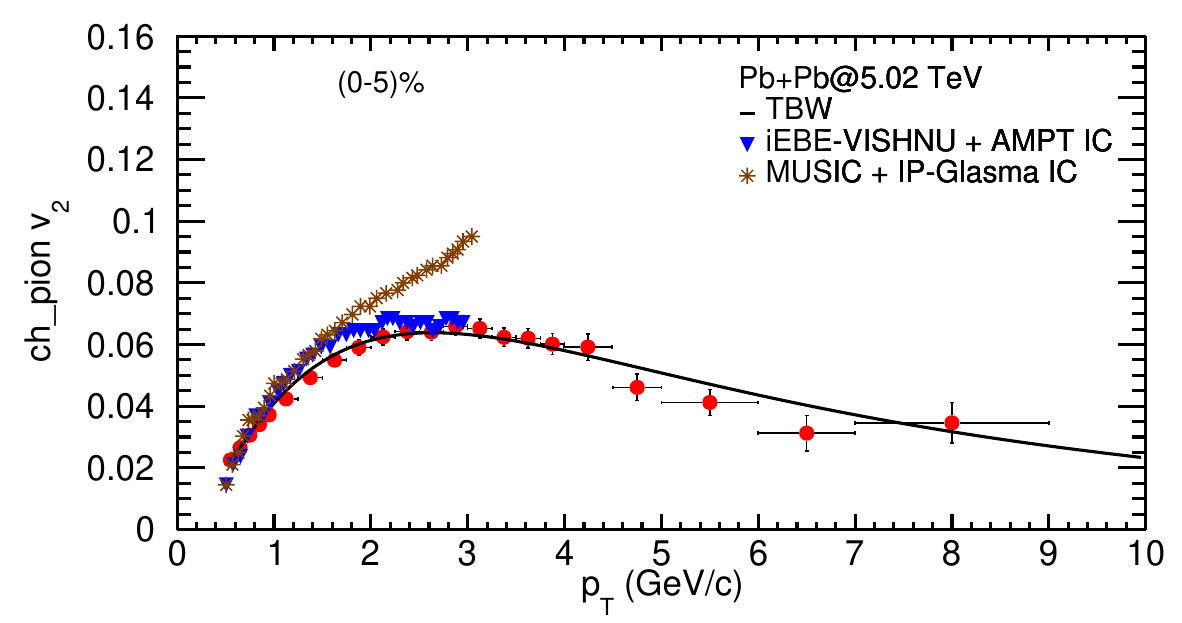}
\end{minipage}}
\hfill
\subfigure[\ $v_3$ of $\pi^{\pm}$ as a function of $p_T$ compared with the hydrodynamical models at $\sqrt{s_{NN}} $= 5.02 TeV for 0-5$\%$ centrality.]{
\label{ } 
\begin{minipage}[b]{0.48\textwidth}
\centering \includegraphics[width=\linewidth]{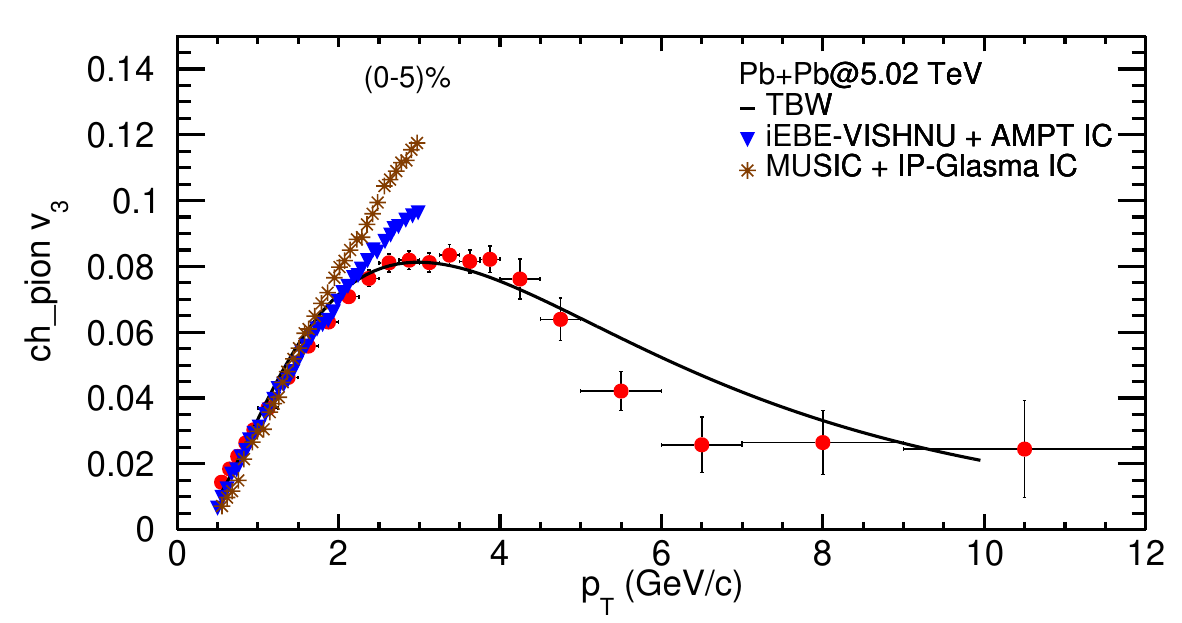}
\end{minipage}}
\hfill
\subfigure[\ $v_4$ of $\pi^{\pm}$ as a function of $p_T$ compared with the hydrodynamical models at $\sqrt{s_{NN}} $= 5.02 TeV for 0-5$\%$ centrality.]{
\label{fig5b} 
\begin{minipage}[b]{0.48\textwidth}
\centering \includegraphics[width=\linewidth]{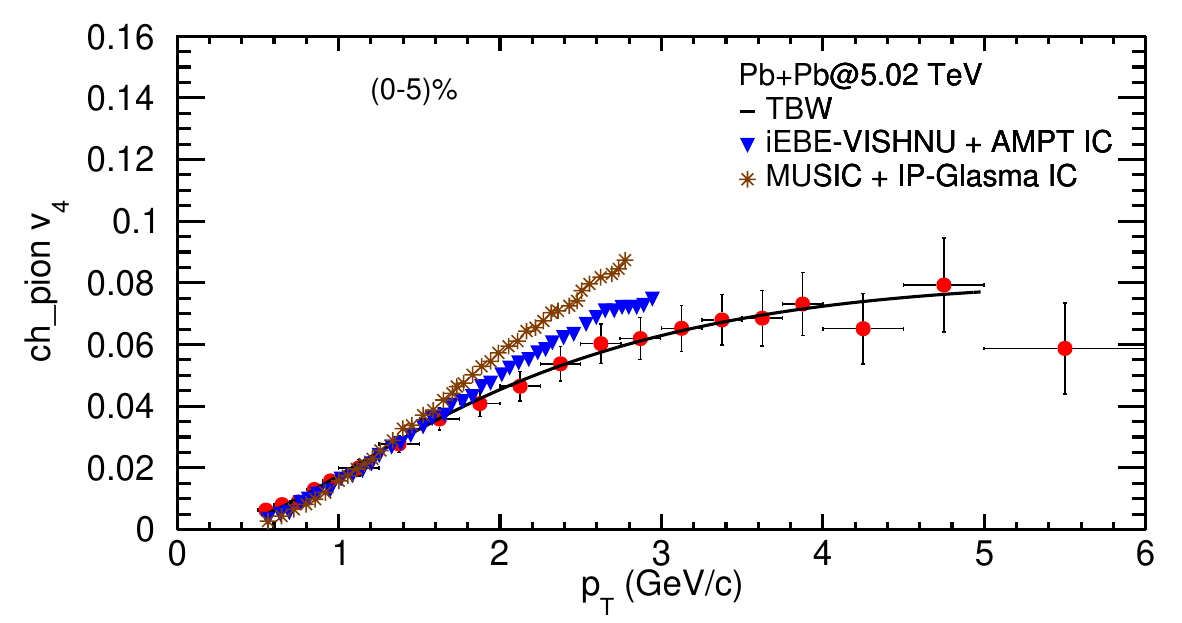}
\end{minipage}}
\caption{Comparison of the higher flow harmonics ($v_2, v_3, v_4$) with the hydrodynamical calculations from MUSIC model using IP-Glasma initial conditions~\cite{ALICE:2018yph,McDonald:2016vlt} and the iEBE-VISHNU hybrid model using AMPT hydrodynamical models~\cite{ALICE:2018yph,Zhao:2017yhj} in Pb-Pb collisions at $\sqrt{s_{NN}} $= 5.02 TeV for the most central collisions.}
\label{hydro} 
\end{figure*}

The azimuthal modulation amplitude ($\rho_a$) is defined as the ratio of the average momentum anisotropy in the transverse plane to the initial spatial anisotropy in the collision zone. In simpler terms, it quantifies the preference of the emitted particles to move in a direction perpendicular to the collision's symmetry plane. As shown in the figures~\ref{fig8} and~\ref{fig9}, $\rho_a$ increases as we go from most central to peripheral collisions. The possible reasons behind this observations are summarized as:

\begin{enumerate}

\item In the peripheral collisions, the medium created in collision experiments may have a shorter lifespan due to the lower energy densities. This shorter duration means that the medium has less time to evolve and can retain the initial anisotropies present in the initial conditions, contributing to a larger $\rho_a$.

\item There are fewer final-state interactions among particles during the late stages of the collision in the peripheral collisions. This reduced number of interactions allows the initial anisotropies to be better preserved in the final particle distributions, resulting in a larger $\rho_a$.

\end{enumerate}
Figure~\ref{hydro} displays the comparison of the $v_2$, $v_3$, and $v_4$ for the charged pions ($\pi^{\pm}$) calculated as a function of $p_T$ in the present formulations with the hydrodynamical calculations from MUSIC model using IP-Glasma initial conditions~\cite{ALICE:2018yph,McDonald:2016vlt} and the iEBE-VISHNU hybrid model using AMPT hydrodynamical models~\cite{ALICE:2018yph,Zhao:2017yhj} in the most central Pb-Pb collisions at $\sqrt{s_{NN}} $= 5.02 TeV. We notice that the iEBE-VISHNU hydrodynamic computations effectively capture the observed azimuthal anisotropy ($v_2, v_3, v_4$) of $\pi^{\pm}$ for $p_T <$ 2.5 GeV/c. Conversely, the MUSIC model aligns with the measurements only upto $p_T < $ 1 GeV/c. Here, we emphasize to highlight that neither of the hydrodynamical models can completely describe the experimental data. The formulations proposed in this work fits the measured experimental data successfully upto $p_T$ = 10 GeV/c for $v_2$ and $v_3$ and upto $p_T$ = 5 GeV/c for $v_4$. However, our approach is completely different from the hydrodynamical calculations.                                                                                                                                                                                                              

\section{Summary and Outlook}
\label{summary}

In summary, we have fitted the $p_T$-spectra, higher flow harmonics ($v_2$, $v_3$, $v_4$) of identified hadrons such as pions, kaons and protons for Pb-Pb and Xe-Xe collisions for LHC energies at various centralities using the BTE in RTA with Tsallis Blast Wave (TBW) function as an equilibrium distribution. The main findings of this analysis are summarized as follows:
 
 \begin{enumerate}
 
 \item The value of $\chi^2/ndf$ is found to be smaller than unity because the point-to-point systematic errors, which are included in the fit and dominate over statistical ones, are estimated on the conservative side and might not be completely random.   
 
 \item The value of the flow profile parameter, $n$ increases from the most central collisions towards peripheral collisions except for proton in Xe-Xe collisions. The large values found in peripheral collisions are maybe due to the spectrum not being thermal over the full range and increases to reproduce the power-law tail.
 
 \item The non-extensive parameters, $q_{pp}$ and $q_{AA}$ increases as we go from the most central towards the peripheral collisions which is expected as these are greater for the system away from equilibrium.
 
 \item The average transverse flow velocity decreases with the mass for both the Pb- Pb and Xe- Xe collisions and decreases from most central to peripheral collisions. Xe-Xe collisions at $\sqrt{s_{NN}}$ = 5.44 TeV has a higher average transverse flow velocity compared to the Pb-Pb collisions at $\sqrt{s_{NN}}$ = 5.02 TeV. Further, the extracted kinetic freeze-out temperature increases from peripheral to most central collisions and decreases with the collisions energies. These findings suggest that the initial higher energy density is responsible for the longer expansion time of the system which results in a larger flow velocity and lower kinetic freeze-out temperature.
 
  \item The parameter $t_f/\tau$ increases with the mass in both Pb-Pb and Xe-Xe collisions while it is not showing any trend centrality wise which needs further investigations and will be discussed in our future work. It suggests that the system has had enough time to reach local thermal equilibrium before the freeze-out occurs. In this case, the observed particle spectra and properties may reflect a system that has experienced substantial equilibration and thermalization.
 
 \item In Pb-Pb and Xe-Xe collisions, the centrality-dependent behaviour of $\rho_a$ for $v_2$,  $v_3$, and $v_4$ exhibits an increase as moving towards the peripheral collision. This trend is consistent with the fact that there are fewer final-state interactions among particles during the late stages of the collision in the peripheral collisions. This reduced number of interactions allows the initial anisotropies to be better preserved in the final particle distributions, resulting in a larger $\rho_a$.

 \end{enumerate} 
 The observation of the centrality-dependent behaviour of $\rho_a$ across both Pb-Pb and Xe-Xe collisions highlights the common underlying physics governing anisotropic flow phenomena. These trends are rooted in the interplay of particle interactions, momentum conservation, and the collective expansion dynamics of the collision systems. The systematic analysis of $\rho_a$ for different flow harmonics enriches our understanding of the collision processes and the role of various particle species in heavy-ion collisions.

\section*{ACKNOWLEDGEMENTS}
SKT acknowledges the financial support of the seed money grant provided by the University of Allahabad, Prayagraj, Uttar Pradesh.

\section*{Data Availability Statement}
The data that support the findings of this study are available upon request from the authors.

\bibliography{v2TBW}

\end{document}